\documentclass[aip,amsmath,amssymb,reprint]{revtex4-2}

\usepackage{graphicx}
\usepackage[utf8]{inputenc}
\usepackage[T1]{fontenc}
\usepackage{etoolbox}
\usepackage{xcolor}
\usepackage{siunitx}
\usepackage{afterpage}
\usepackage{capt-of}
\usepackage{needspace}

\makeatletter
\def\@email#1#2{%
  \endgroup
  \patchcmd{\titleblock@produce}
    {\frontmatter@RRAPformat}
    {\frontmatter@RRAPformat{\produce@RRAP{*#1\href{mailto:#2}{#2}}}\frontmatter@RRAPformat}
    {}{}
}%
\makeatother

\begin{document}

\preprint{AIP/123-QED}

\title[A cryogenic apparatus for coupling two-dimensional materials to a confocal multimode optical cavity]%
      {A cryogenic apparatus for coupling two-dimensional materials to a confocal multimode optical cavity}

\author{Han S.~Hiller}
\affiliation{Department of Physics, Stanford University, Stanford, CA 94305, USA}
\affiliation{E.~L.\ Ginzton Laboratory, Stanford University, Stanford, CA 94305, USA}
\author{Pranav Parakh}
\affiliation{E.~L.\ Ginzton Laboratory, Stanford University, Stanford, CA 94305, USA}
\affiliation{Department of Applied Physics, Stanford University, Stanford, CA 94305, USA}
\author{Samuel H.~Aronson}
\affiliation{Department of Physics, Stanford University, Stanford, CA 94305, USA}
\affiliation{E.~L.\ Ginzton Laboratory, Stanford University, Stanford, CA 94305, USA}
\affiliation{Department of Applied Physics, Stanford University, Stanford, CA 94305, USA}
\author{Kenji Maeda}
\affiliation{Department of Physics, Stanford University, Stanford, CA 94305, USA}
\affiliation{E.~L.\ Ginzton Laboratory, Stanford University, Stanford, CA 94305, USA}
\author{Di Lao}
\affiliation{E.~L.\ Ginzton Laboratory, Stanford University, Stanford, CA 94305, USA}
\author{Julian Stewart}\affiliation{Department of Physics, University of Washington, Seattle, Washington 98195, USA}
\affiliation{Department of Materials Science and Engineering, University of Washington, Seattle, Washington 98195, USA}
\author{Zengde She}\affiliation{Department of Physics, University of Washington, Seattle, Washington 98195, USA}
\affiliation{Department of Materials Science and Engineering, University of Washington, Seattle, Washington 98195, USA}
\author{Jierong Wang}\affiliation{E.~L.\ Ginzton Laboratory, Stanford University, Stanford, CA 94305, USA}
\affiliation{Department of Applied Physics, Stanford University, Stanford, CA 94305, USA}
\author{Xiaodong Xu}\affiliation{Department of Physics, University of Washington, Seattle, Washington 98195, USA}
\affiliation{Department of Materials Science and Engineering, University of Washington, Seattle, Washington 98195, USA}
\author{Tony Heinz}
\affiliation{E.~L.\ Ginzton Laboratory, Stanford University, Stanford, CA 94305, USA}
\affiliation{Department of Applied Physics, Stanford University, Stanford, CA 94305, USA}
\affiliation{SLAC National Accelerator Laboratory, Menlo Park, California 94025, USA}
\author{Benjamin L.~Lev}
  \email{benlev@stanford.edu}
\affiliation{Department of Physics, Stanford University, Stanford, CA 94305, USA}
\affiliation{E.~L.\ Ginzton Laboratory, Stanford University, Stanford, CA 94305, USA}
\affiliation{Department of Applied Physics, Stanford University, Stanford, CA 94305, USA}

\date{\today}

\begin{abstract}
Two-dimensional van der Waals materials exhibit a variety of correlated electron phases, and optical driving offers a promising route toward manipulating them. For example, cavity-enhanced, continuous-wave (CW) Raman excitation has been suggested as a way to coherently and superradiantly populate phonons or charge density waves via material excitons. A steady-state phonon population may be sustained with sufficiently strong electron-phonon coupling to drive novel collective response. We describe an apparatus built to meet the requirements of such an experimental program: Namely, an ultrahigh-vacuum system housing a length-tunable confocal Fabry--P\'{e}rot cavity with an intracavity sample, both cryogenically cooled and stabilized against vibrations. A four-axis nanopositioner aligns the sample and supports electrical leads for sample carrier density modulation and transport measurements. Transmission through the multimode cavity enables in situ sample imaging for alignment; the sample is a transition metal dichalcogenide in this work. Operating near the confocal geometry concentrates the optical field into a localized supermode that substantially enhances light–matter coupling.  This enhancement is preserved despite the millimeter-scale cavity length, which provides room for sample alignment and exchange. 
\end{abstract}

\maketitle
\thispagestyle{plain}

\section{\label{sec:intro}Introduction}

Van der Waals (vdW) heterostructures are a class of quantum materials that have emerged as promising testbeds for correlated electron physics and as potential devices for quantum technologies~\cite{Geim2013}.  A remarkable number of interesting states of matter have been observed in these atomically thin, two-dimensional (2D) systems, including unconventional superconductivity~\cite{Cao2018, Xi2016}, correlated insulating states~\cite{Lu2019}, topological phenomena~\cite{Fei2017, Sajadi2018}, and novel forms of magnetism~\cite{Sharpe2019,Huang2017}. The microscopic mechanisms for some of these behaviors remain unclear~\cite{Mak2022smm}. 

Intentionally driving phonon modes may be a route toward enhancing electronic properties in these systems, e.g., by raising the superconducting critical temperature~\cite{Fausti2011lsi,Mankowsky2014nld,Mitrano2016pls,Cavalleri2017ps}.  However, current schemes use pulsed THz or mid-infrared lasers to directly drive these modes, and the transient nature of this drive complicates measurement interpretation~\cite{Disa2021ecs}. Short pulses obscure the steady-state response, because excitation and relaxation dynamics are typically incompatible with the slow timescales of transport measurements. By contrast, light-matter coupling using continuous-wave (CW) optical lasers offers an exciting new route: By driving a steady-state phonon population rather than a transient one, CW stimulation of material excitations is more naturally compatible with traditional probes like electrical transport as a diagnostic of the resulting electronic phase.

Monolayer transition metal dichalcogenides (TMDs) are a particularly compelling subclass of vdW materials for exploring the efficacy of CW optical-field driving~\cite{Bourzutschky2024rci}. As monolayers, these semiconductors become direct-gap and host tightly bound excitons---electron-hole pairs with binding energies exceeding $\sim$500\,meV---whose optical transitions lie in the visible range~\cite{Chernikov2014}. Encapsulation in hexagonal boron nitride (hBN) dramatically narrows the exciton linewidth, enabling coherent optical access to the valley degree of freedom at the $K$ and $K'$ points of the Brillouin zone~\cite{Cadiz2017}. These spectrally sharp, optically active excitons behave in a manner somewhat akin to atomic resonances, admitting light-matter coupling schemes known in atom-based cavity quantum electrodynamics (QED)~\cite{Dimer2007pro,Baumann2010dqp,Kroeze2018sso}. But driving excitons also allows us to couple strongly to Raman-active phonon modes~\cite{Bourzutschky2024rci}. This property makes TMDs an ideal material platform for driving phonons via their interaction with exciton-polaritons, the quasiparticles formed from strongly mixing excitons and photons via cavity QED.  The result would be CW, light-mediated, phonon stimulation and control.  How these coherent phonons would then drive electronic behavior is a subject of active study that we aim to address with the experimental system presented here, benchmarking future theoretical work.

Using the field of a cavity can greatly increase the coupling of 2D materials to light for the purpose of affecting material properties~\cite{Carusotto2013qfo,Sanvitto2016trt,Schlawin2019ces,Curtis2019cqe,Gao2020pep,Curtis2022cmi,Schlawin2022cqm,Bretscher2026fei}.
In Ref.~\cite{Bourzutschky2024rci}, we proposed a scheme to drive phonons in monolayer TMDs via a cavity-enhanced (two-photon) Raman transition. An off-resonant pump laser (incident upon the material at an angle transverse to the cavity axis)  drives material exciton-phonon modes as photons are Rayleigh scattered into the optical cavity. Initially in the vacuum state, the cavity mode becomes populated and acts to depump the exciton-phonon excitation into a purely phononic mode. Thus, the cavity mediates a two-photon transition from the phonon-free ground state, via a virtually excited exciton, to a coherently driven phonon state.

The cavity plays a dual role. First, it enhances the electric field, enabling the Raman transition with only a few photons and thereby minimizing optically induced heating. Second, the cavity field permits a Hepp–Lieb–Dicke superradiant phase transition~\cite{Kirton2018sal} above a critical pump strength, generating a steady-state coherent phonon population that grows linearly with the number of driven excitons $N$, with collective coupling enhanced by the standard Dicke factor $\sqrt{N}$. This macroscopic phonon amplitude can substantially enhance effective electron–phonon coupling in the material.

Planar microcavities---where Bragg mirrors are grown or deposited directly above and below the material---have demonstrated strong light–matter coupling~\cite{Kimble1998sio} in 2D systems~\cite{Liu2015, Dufferwiel2015,Zhang2021vdw,Vadia2021oic,Drawer2025twm,Hoang2026acr}. In these planar Fabry--P\'{e}rot cavities, the mirrors are either attached to the 2D material or are only microns away, restricting all light to be perpendicular to the sample plane and cavity coupled. Unfortunately, this blocks transverse optical access, directly to the material.  Such access is needed to deliver the transverse pump field required for implementing the aforementioned Raman scheme. By contrast, a more open cavity, with mirrors separated by several millimeters, supports both longitudinal coupling to the cavity mode and independent transverse access. Such a geometry enables direct optical pumping of the material while accommodating auxiliary spectroscopic probes and the electrical contacts necessary for gating and 4-point transport measurement. Moreover, the sample can be oriented at 45 degrees with respect to both the cavity axis and the pump field, optimizing the scattering rate from the transverse pump, off the material, and into the cavity.  

The traditional downside of long cavities is their larger mode volume. That reduces the single-photon field strength, resulting in lower single-emitter cooperativity. However, our group has demonstrated a technique that obviates this. Using a multimode Fabry--P\'{e}rot cavity---such as one operating in the confocal geometry whereby the mirror spacing $L$ equals the mirror radii of curvature $R$---causes many transverse modes to become frequency degenerate~\cite{Vaidya2018tpa,Guo2019spa,Guo2019eab}.  Consequently, the coherent superposition of thousands of cavity modes concentrates the electric field into a cavity-diffraction-limited spot size that is far smaller than the single-mode waist of the cavity.  Indeed, in confocal cavities intended for ultracold atom research, the resulting ``supermode" waist~\cite{Vaidya2018tpa} of an $L=1$~cm cavity is as small as 1.7~$\mu$m, far narrower than the 35-$\mu$m Gaussian mode waist~\cite{Kroeze2023hcu}.  This resulted in a (dispersive) single-atom cooperativity for virtual photon processes of 110, far larger than the single-mode, single-atom cooperativity of 5. By recovering much of the cooperativity advantage of planar microcavities, we expect the interaction of TMD excitons and confocal cavity light to reach the high-cooperativity regime~\cite{Bourzutschky2024rci}.

Building on this framework, we recently proposed in Skolc \textit{et al.}~\cite{Skolc2026scd} that a related confocal cavity-driven scheme using doped TMDs can induce superradiant charge density waves---a direct cavity-driven electronic phase transition---further demonstrating the potential of CW optical-cavity-coupling as a driver of correlated electronic physics. 

A useful byproduct of operating the cavity in a confocal multimode configuration is that it functions as an in situ imaging system for the sample. The degenerate transverse modes map the sample plane onto the output mode, so that illumination injected through the input mirror produces, upon transmission, an image in which sample absorption appears as spatial contrast on the cavity emission. We use this capability to register the TMD flake to the cavity mode center.  The registration precision is set by the aforementioned supermode waist of the cavity.  Although the technique is employed here primarily as an alignment diagnostic rather than as a measurement, it eliminates the need for an auxiliary imaging objective.  Any
cryogenic cavity-sample experiment where integrating an alternative imaging path is inconvenient may benefit from multimode cavity imaging. We now describe an apparatus designed to meet these requirements.

\section{\label{sec:overview}Apparatus Overview}

\begin{table}
\caption{Performance specifications of the apparatus.
All values refer to steady-state operation unless noted. Mode waists, $w_0$, refer to the Gaussian radius of the intensity.}
\begin{tabular}{lp{3.5cm}}
\hline\hline
Parameter & Value \\
\hline
\multicolumn{2}{l}{\textit{Sample environment}} \\
Sample base temperature       & 10.4~K \\
Heat shield temperatures       & 100--110~K \\
Sample exchange \& pump-down time  & 1~day \\
Cooldown time (300\,K $\to$ base) & 10~hrs \\
Sample nanopositioner range      & $\pm$2.5~mm \\
Sample positioning resolution & 100~nm  \\
\hline
\multicolumn{2}{l}{\textit{Vacuum}} \\
Base pressure (cold)          & $10^{-10}$~Torr  \\
Base pressure (warm, post-bake) & $10^{-8}$~Torr \\
\hline
\multicolumn{2}{l}{\textit{Optical cavity}} \\
Mirror radius of curvature    & $R= 10.00(0.05)$~mm \\
Mirror reflectivity           & upper: 99.7\%, lower: 97.7\% \\
Mirrors coated for            & 600--900\,nm  \\
Cavity length                 & $L=R$ \\
Finesse                       & 208 (181 w/substrate) \\
Linewidth (FWHM)              & 64~MHz (83\,MHz w/substrate) \\
Mode waist $w_0$ (TEM$_{00}$) & 35~$\mu$m  \\
Supermode waist $\xi$, test cavity         & 5~$\mu$m \\
Supermode waist $\xi$, science cavity         & 9.5~$\mu$m  \\
\hline
\multicolumn{2}{l}{\textit{In situ imaging \& alignment}} \\
Imaging field of view         & 500~$\mu$m  \\
Resolution & $\sim$4~$\mu$m \\
\hline
\multicolumn{2}{l}{\textit{Mechanical \& thermal stability}} \\
RMS cavity length noise (cryo off) & 97(9)~pm \\
RMS cavity length noise (cryo on)  & 110(10)~pm  \\
Sample vibrational motion (cryo on) & <8~nm \\
\hline\hline
\end{tabular} \label{specstable}
\end{table}

\afterpage{
\begin{figure*}
  \includegraphics[width=\textwidth]{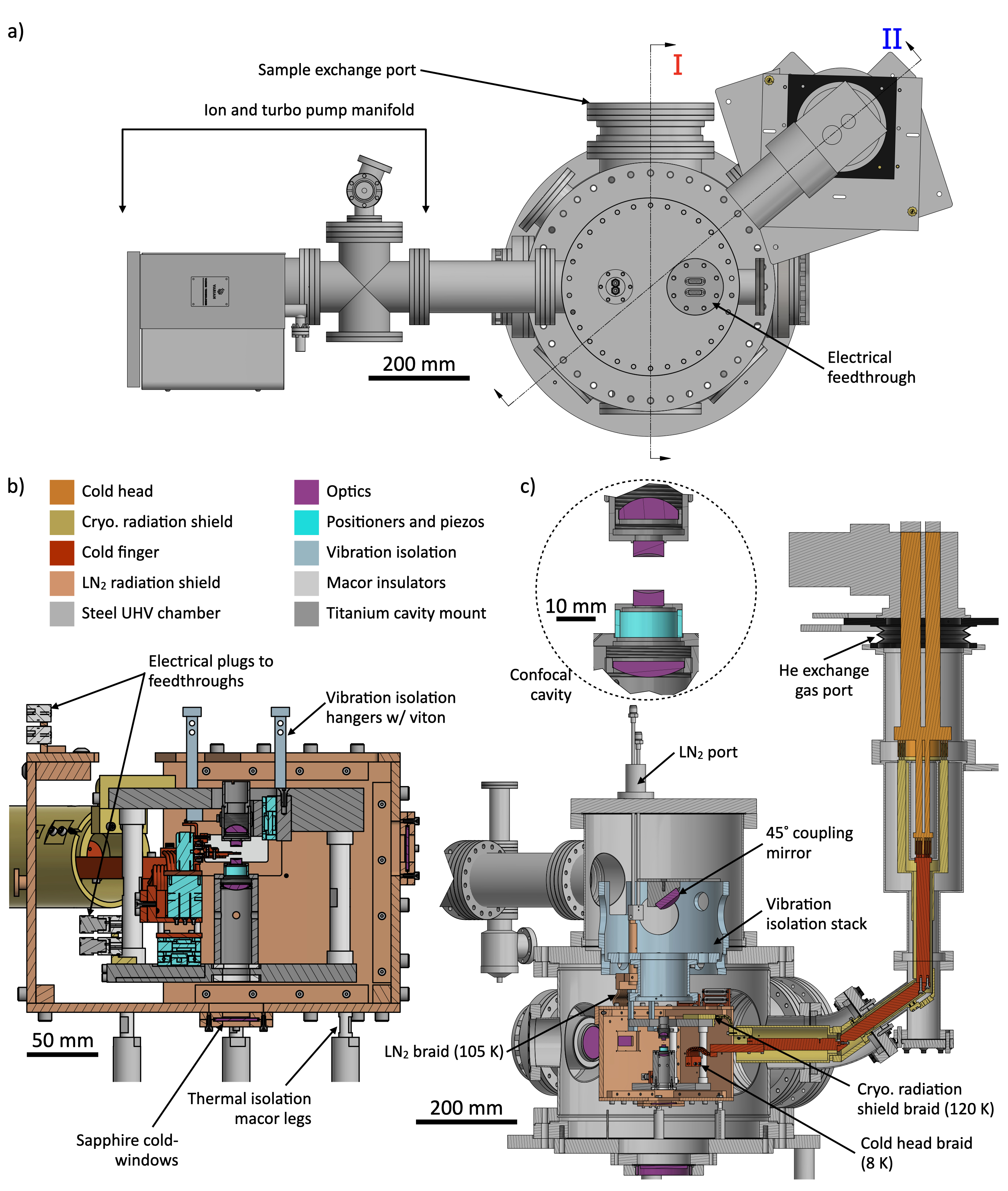}
  \caption{Overview of the apparatus. (a) Top view of the full vacuum assembly.  The sample exchange port is a 10" conflat window that is removed during exchange.  (b) Cross-sectional view of the assembly, sliced along line I in panel (a), which is the centerline of the cavity mount and sample stack. Important components are labeled.  (c) Cross-sectional view of the assembly, sliced along line II, the centerline of the cryostat. The sample stage has been removed for clarity, though the copper cold finger braid is shown. (Inset) Magnified view of the cavity and coupling lenses.}
  \label{fig:apparatus_overview}
\end{figure*}
}

\medskip
The system enables the coupling of a confocal Fabry--P\'{e}rot cavity to an optically driven, cryogenically cooled 2D sample. We summarize the system capabilities in Table~\ref{specstable}.
The cavity and sample are mounted in  an ultrahigh vacuum (UHV) chamber; see Fig.~\ref{fig:apparatus_overview} for a sketch of the apparatus. The chamber is maintained at UHV pressures to reduce sample contamination and to allow measurements at room temperature in the absence of cryopumping. Vibrations from the pulse-tube cryocooler are mitigated by decoupling the cooler from the cavity; we use techniques developed in Refs.~\cite{Naides2013tug,Kollar2015aac,Taylor2021asq}. (Section \ref{sec:cryo} details the cryogenic and vacuum systems.)
The vibration isolation allows the cavity length to be stabilized at the confocality point as well as the sample to be held at an antinode of the cavity field. The cavity mirrors carry a broadband dielectric coating spanning relevant deep-red exciton resonances in the TMDs we employ, MoSe$_2$ and WSe$_2$. See Section \ref{sec:cavity_design} for details.

The cavity mirrors and the sample are both optically accessible via chamber windows, similar to the design in Ref.~\cite{Taylor2021asq}. Optical access is provided through sapphire windows thermally anchored to the cryogenic radiation shields, reducing blackbody radiation incident on the sample.  The confocal cavity itself allows optical imaging of the sample in situ owing to its multimode nature; see Section \ref{sec:cavity_modes} for details. The chamber provides electrical feedthrough access for in situ sample carrier density modulation, transport measurements, and thermometry.

The sample is attached to a thin, transparent, antireflection-coated sapphire substrate. The cavity mode propagates through the sample and substrate unobstructed by the copper ``stick" that supports it.  This stick extends between the cavity mirrors and hosts the electrical leads for transport measurements and a temperature probe. A four-axis nanopositioner provides in situ placement and rotational-angle control of the 2D sample within the cavity mode; see Section \ref{sec:sample} for details. 

The copper stick is cryogenically cooled with a closed-cycle cryostat, providing a sample base temperature of $10.4$~K.  The cavity mirrors are cooled to provide radiation shielding for the sample mounted in between.  The vibrationally isolated cold finger and its heat shield are designed to enter the UHV chamber from an angle that does not occlude optical access to the cavity mirrors or sample. Cooling for the heat shield around the sample (not covered by the mirror) is provided by both the cold finger heat shield and thermal contact to a separate LN$_2$ flow cryostat.

\section{\label{sec:sample}Sample Stage}

We now describe the manner in which the sample is mounted and manipulated within the chamber.

\subsection{\label{sec:mount}Sample Mount Design}

The sample is mounted on a transparent sapphire substrate that is positioned between the cavity mirrors. The substrate is coated with an antireflective coating that maintains high optical transmission at the relevant wavelengths.  This eliminates any apparent intracavity etaloning (except when the substrate is normal to the cavity axis---not a configuration in which we operate). Sapphire is chosen for its high thermal conductivity at cryogenic temperatures and its optical transparency over the spectral range used in the experiment.  The sample, typically an hBN-encapsulated TMD, is attached to one surface of the substrate during the sample fabrication process.

The substrate is fixed with thermally conductive adhesive to the copper sample holder to provide a thermal link to the cryostat cold stage. Thermal contact is provided by flexible oxygen-free copper braids that connect the sample mount to the cryostat cold finger, allowing efficient heat removal while minimizing mechanical coupling to the cryostat. 

\subsection{\label{sec:motion}4-Axis Sample Motion}
\begin{figure}
  \centering
  \includegraphics[width=\columnwidth]{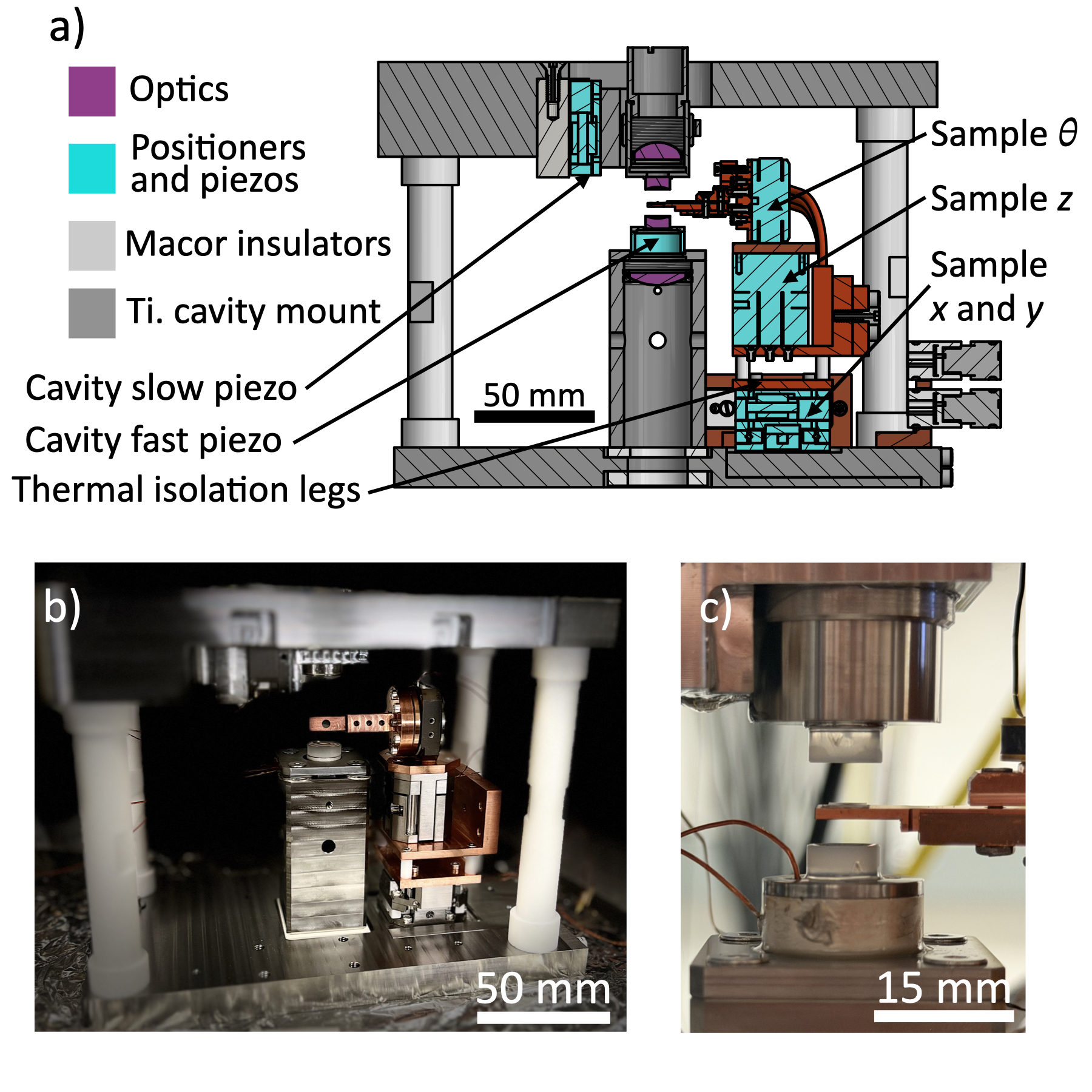}
  \caption{Cavity mount and sample positioning stage. 
    (a) Schematic of the cavity mount, showing in purple (from top to bottom) the coupling lens, cavity mirrors,  and collimating lens.  In cyan are the sample stack attocubes, the cavity attocube, and the fast piezo. (b) Photograph of the cavity mount, before installation into the vacuum chamber. (c) Photograph of the cavity in the chamber with a 2D sample positioned at the cavity mid-plane.}
  \label{fig:cavity mount}
\end{figure}

Precise positioning of the sample relative to the cavity mode is achieved using a four-axis cryogenic nanopositioning stage depicted in Fig.~\ref{fig:cavity mount}. Motion is implemented using UHV and cryogenic-compatible slip–stick piezoelectric actuators. The stage provides translation along the $x$, $y$, and $z$ axes as well as rotation perpendicular to the cavity axis. Starting from the sample, the copper stick is mounted on the rotation stage, which itself is oriented at 90$^\circ$ with respect to the $z$ stage upon which it is mounted.  These are attached to the $x$ and $y$ stages, which are fixed to the titanium block that forms the base of the sample stage (see Fig.~\ref{fig:cavity mount}a for details). Titanium is used for the sample stage and the cavity mount for its high stiffness. This increases the vibrational frequency of mechanical resonances into the vibration-isolation suppression band of the mass-spring support structure. 

The vibrational low-pass-filtering structure, shown in blue in Fig.~\ref{fig:apparatus_overview}c, is comprised of two heavy steel cylinders.  Each is hung off a higher platform by Viton rubber beads that serve as springs.  At the bottom, the sample stage hangs from 4 titanium legs off the bottom cylinder, again with Viton spacers. Together, this forms a stiff three-stage, vibration-isolation stack. 

The translational degrees of freedom of the sample stage allow it to be positioned at the antinode of the cavity standing wave and centered within the cavity supermode. Electrical connections to the piezo actuators are routed through UHV-compatible wiring and vacuum feedthroughs. The rotational degree of freedom enables alignment of the sample normal relative to the cavity axis, $\theta$.  Rotation allows us to transversely pump the sample at various angles. We observe sidebands on the cavity transmission when the sapphire substrate is normal to the cavity axis (i.e., $\theta = 0$).  This is due to an etaloning effect, but fortunately our experiments will require  $\theta \approx 45^\circ$ to the cavity axis, and we do not observe etaloning beyond a $\theta$ of a few degrees.

The modular design of the stage allows samples to be exchanged without modification of the surrounding cavity structure. We can typically remove the copper stick without removing any of the translation stages or the rotation stage.  Samples attached to different sticks are easily swapped in once the main 10'' chamber window is removed and a panel of the heat shield is unscrewed.   A full sample exchange and pump-down cycle can typically be completed within one day without a chamber bakeout, enabling a rapid iteration sequence for device characterization.

\section{\label{sec:cavity_sec}The Multimode Cavity}

We now describe the mirrors and length-tunability of the cavity. 

\subsection{\label{sec:cavity_design}Design and Mirror Specifications}
\begin{figure*}
  \centering
  \includegraphics[width=\textwidth]{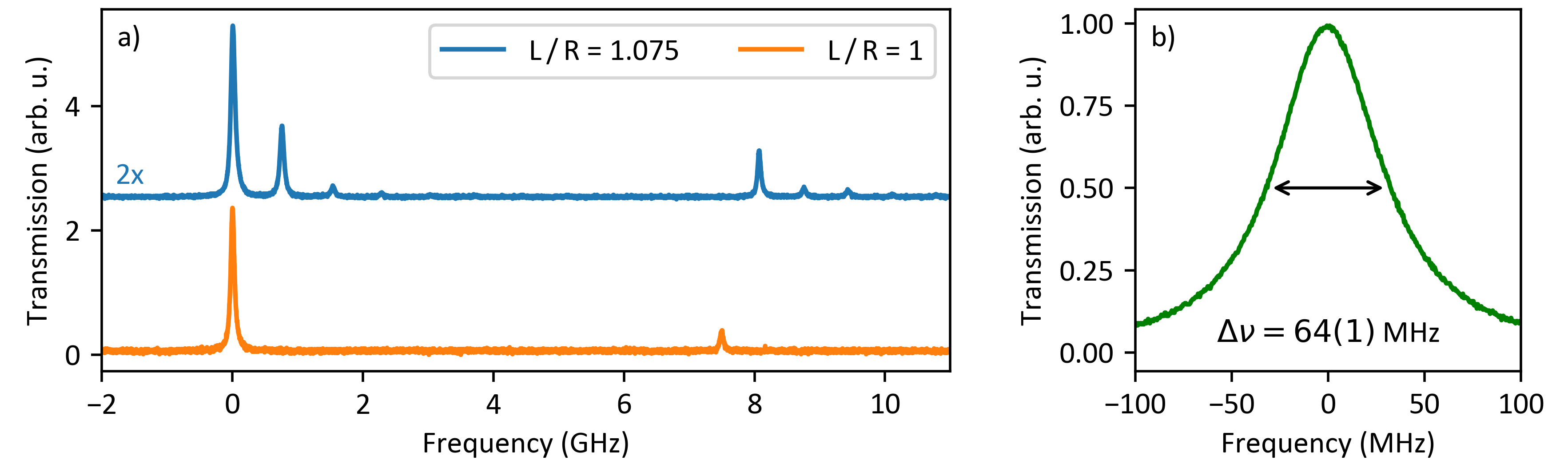}
  \caption{Cavity spectrum and linewidth. (a) Transmission spectra of near-confocal (blue curve) and confocal (orange curve) optical cavity geometries. The near-confocal curve is scaled vertically by $2\times$ and offset for clarity. (b) High-resolution scan over the TEM$_{00}$ transmission peak taken away from confocality at $L/R = 1.125$, exhibiting a linewidth of 64(1) MHz.}
  \label{fig:cavity_spectrum}
\end{figure*}

\begin{figure*}[t!]
    \includegraphics[width=\textwidth]{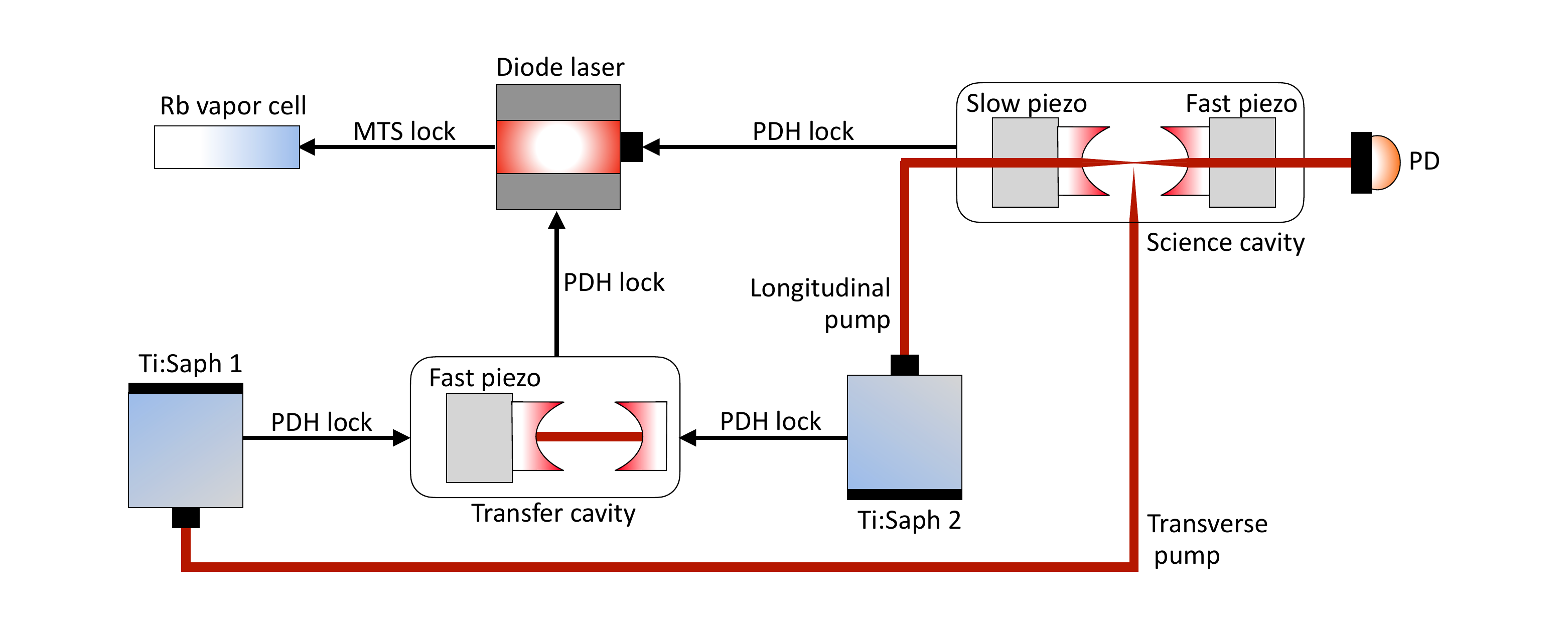}
    \caption{Schematic of the laser and cavity stabilization scheme. The arrows point toward the device to which the element is being locked.  1) The diode laser (DL) is stabilized to a Rb vapor cell via a modulation transfer spectroscopy lock. 2) The science cavity length is locked to the DL via a Pound-Drever-Hall (PDH) lock, with the transmitted signal detected on a photodiode (PD). 3) The transfer cavity is locked to the DL via PDH. 4) The two titanium-sapphire lasers (Ti:Saphs) are independently stabilized to the transfer cavity via PDH locks at any chosen wavelengths. The Ti:Saphs serve as the longitudinal and transverse pump beams for the science cavity and sample, respectively.}
    \label{fig:locking}
\end{figure*}

The sample is positioned within a tunable-length Fabry--P\'{e}rot cavity formed by two opposing mirrors with radius of curvature $R = 1$~cm. At confocality $L=R$, the free spectral range (FSR) is approximately $15$~GHz. The mirrors are fabricated from superpolished fused-silica substrates and are coated for high reflectivity in the $600$--$900$~nm range. (The flat back-surfaces are antireflection-coated.)
For the bare cavity (no substrate or sample inside), we measure a finesse of $\mathcal{F} = 208$ for the TEM$_{00}$ mode when the cavity length is far from the confocal point. That is,  the linewidth of the TEM$_{00}$ transmission peak at $L/R = 1.125$ is $\Delta \nu =$ 64(1)~MHz; see Fig.~\ref{fig:cavity_spectrum}b.

The cavity length is tunable in situ using a slip-stick piezoelectric actuator on the top mirror.  The travel is over several millimeters which allows us to precisely tune the cavity to the $L=R$ confocal degeneracy point or operate single mode for benchmarking.  Fine stabilization of the cavity length, i.e., locking this cavity resonance to a frequency stabilized diode laser, is achieved using a fast piezo actuator attached to the bottom mirror. (The diode laser is itself locked to a Rb vapor cell at 780~nm using modulation-transfer spectroscopy~\cite{Shirley1982}.)  An error signal is derived using the Pound-Drever-Hall (PDH) technique~\cite{Black2001ait}, and fed back to the fast piezo.

Phase stability between the longitudinal and transverse pump lasers and the cavity is essential for the coherent driving of the intracavity field. Our stabilization scheme is shown in Fig.~\ref{fig:locking}. The two titanium-sapphire pump lasers, which are tunable from ${\sim}600-800$~nm, are referenced, via an external ``transfer" cavity,  to the same diode laser used to lock the ``science" cavity. (We designate the cavity within the vacuum chamber as the science cavity.)  This locking chain ensures that all lasers and the science cavity share a common frequency and phase reference.

\subsection{\label{sec:cavity_modes}Multimode Structure and Supermodes}
When operated at the confocal geometry, the cavity supports a large number of transverse electromagnetic (TEM) modes whose frequencies become degenerate. In an ideal confocal cavity, all even or odd modes are exactly degenerate. In practice, mirror aberrations lift this degeneracy, though many thousands of modes can remain degenerate or nearly so, even in a cavity of finesse a hundred-times-greater than those we require~\cite{Kollar2015aac,Kroeze2023hcu}. Finesse of only a hundred or so is necessary for the aforementioned goals, because the materials themselves are absorptive at the percent level.

\begin{figure*}
  \centering
  \includegraphics[width=\textwidth]{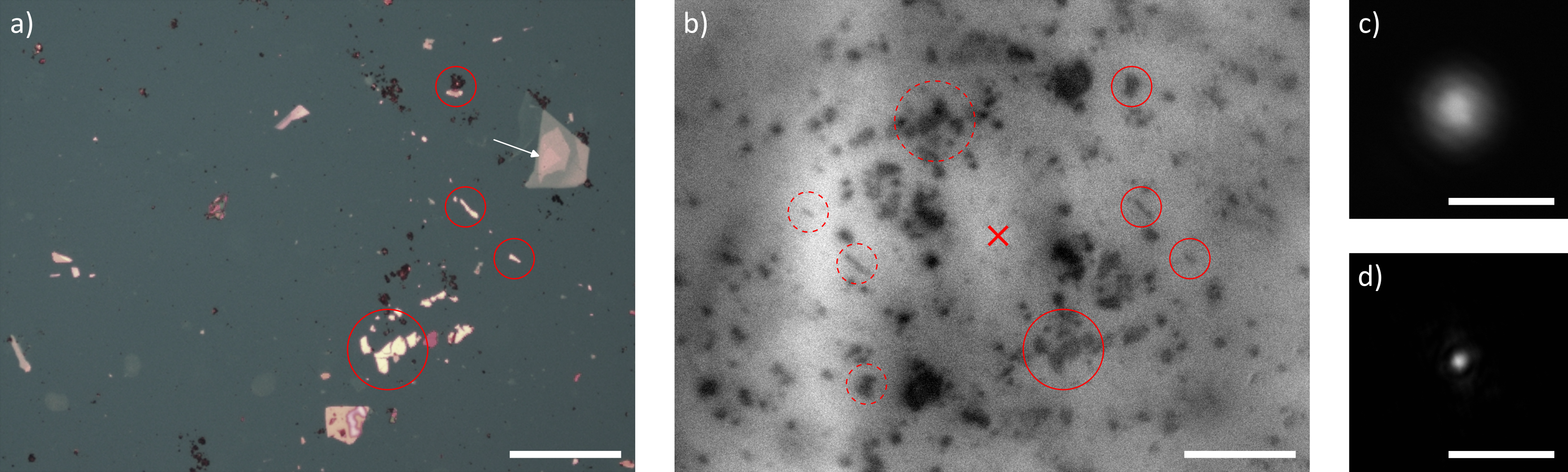}
  \caption{In situ optical imaging of TMD sample. (a) Optical micrograph of the sample taken outside the vacuum chamber under 20$\times$ magnification. Red circles indicate features away from the sample used to align the sample to the cavity supermode. The reader will notice the many errant flakes and fabrication residue on the sapphire substrate. That these do not affect the cavity finesse is because the supermode forms at the transverse location where we will want to pump the cavity, which is at the sample (hBN-encapsulated monolayer WSe$_2$) indicated by the white arrow. (b) Image of the sample taken through the confocal cavity emission when a  broadband LED source is resonantly coupled to cavity modes. The solid red circles indicate the same features as in (a).   Because a confocal cavity resonates on modes of all the same parity, the left half of the image is a copy of the right and formed by rotating 180$^\circ$ about the cavity axis (denoted by the red $\times$). The dashed circles show the virtual images of the features indicated by solid circles. This image has been contrast-enhanced and its background has been reduced by removing low-spatial frequency features, but it has been neither smoothed nor sharpened. (c) The TEM$_{00}$ spot without the substrate and with $L$ tuned away from the confocality point for single-mode operation. The measured waist is 35~$\mu$m. (d) Confocal supermode spot without the intracavity substrate. The measured cavity waist is 9.5~$\mu$m. All scale bars have a length of 100~$\mu$m.}
  \label{fig:sample_imaging}
\end{figure*}

Near to, but not at confocality, the spectrum exhibits a series of closely spaced transverse modes. As the cavity approaches the confocal condition, these modes merge, enabling the formation of a ``supermode"~\cite{Kroeze2023hcu}. The supermode can be understood as a coherent superposition of many transverse cavity modes whose relative phases are determined by the spatial overlap with the sample or a longitudinal pump field (i.e., one that couples directly through the mirrors). As a result, the effective optical mode becomes localized around the position of the material, allowing for a much smaller spot size than that of the single-mode, TEM$_{00}$ mode, as can be seen in Fig.~\ref{fig:sample_imaging}c. We choose to place the sample in the center of the cavity so that the supermode that couples to it is centrosymmetric with respect to the cavity axis. The spatial profile of the resulting supermode is characterized by an effective waist $\xi \approx 9.5~\mu$m, which is significantly smaller than the waist of the fundamental $\mathrm{TEM}_{00}$ mode, 35~$\mu$m.  (A spatial light modulator can improve the mode matching, allowing us to reduce the supermode waist down to 5~$\mu$m like in the test cavity we discuss below.) The properties of the multimode structure are influenced by mirror aberrations, finite mirror size, and residual absorption in the coatings, all of which can lift degeneracies and introduce mode-dependent losses. These effects set the practical limit on the number of modes contributing to the supermode and therefore determine the achievable spatial localization and enhancement of the light–matter interaction.  Nevertheless, this localization enhances the light–matter interaction strength by concentrating the electromagnetic field at the position of the sample. This enhancement is the principal advantage of operating at confocality and directly relaxes the  single-mode coupling strength required to reach the phonon-polariton condensation threshold~\cite{Bourzutschky2024rci}.  In turn, this allows us to employ cavities with mirrors not attached to the material, but rather half-a-centimeter away.

We now compare the transmission spectrum of the near-confocal and confocal regimes in Fig.~\ref{fig:cavity_spectrum}.
Figure~\ref{fig:cavity_spectrum}a shows the near confocal spectrum, where many modes can be easily resolved, as well as the degenerate confocal spectrum with most modes resonating together at the same frequency.  Inserting the sapphire substrate reduces this mode's finesse to $\mathcal{F} \approx 181$, corresponding to a linewidth of $\kappa/2\pi \approx 83$~MHz.  The reduction is likely due to substrate absorption and scattering. 

We next provide finesse measurements for a WSe$_2$ sample on a sapphire substrate with gold gating leads attached.  We measure it in a test cavity setup that is identical to the in-chamber cavity but resides in free space for easy access.   Two parallel gold wires, each of width 3~$\mu$m and separated by 1~$\mu$m, run from the substrate edge to the sample. One wire contacts the backgate and the other serves as a ground for the TMD sample; see Fig.~\ref{fig:sample_with_leads}a.  (Not visible is a thin graphite contact.) This highlights the ability of our open cavity geometry to permit electrical leads to reach nearly to the center of the cavity midplane, thereby enabling gate control and (eventually) transport measurements while preserving optical access along both the longitudinal and transverse directions. 

\afterpage{
\begin{figure}[h!]
    \includegraphics[width=\columnwidth]{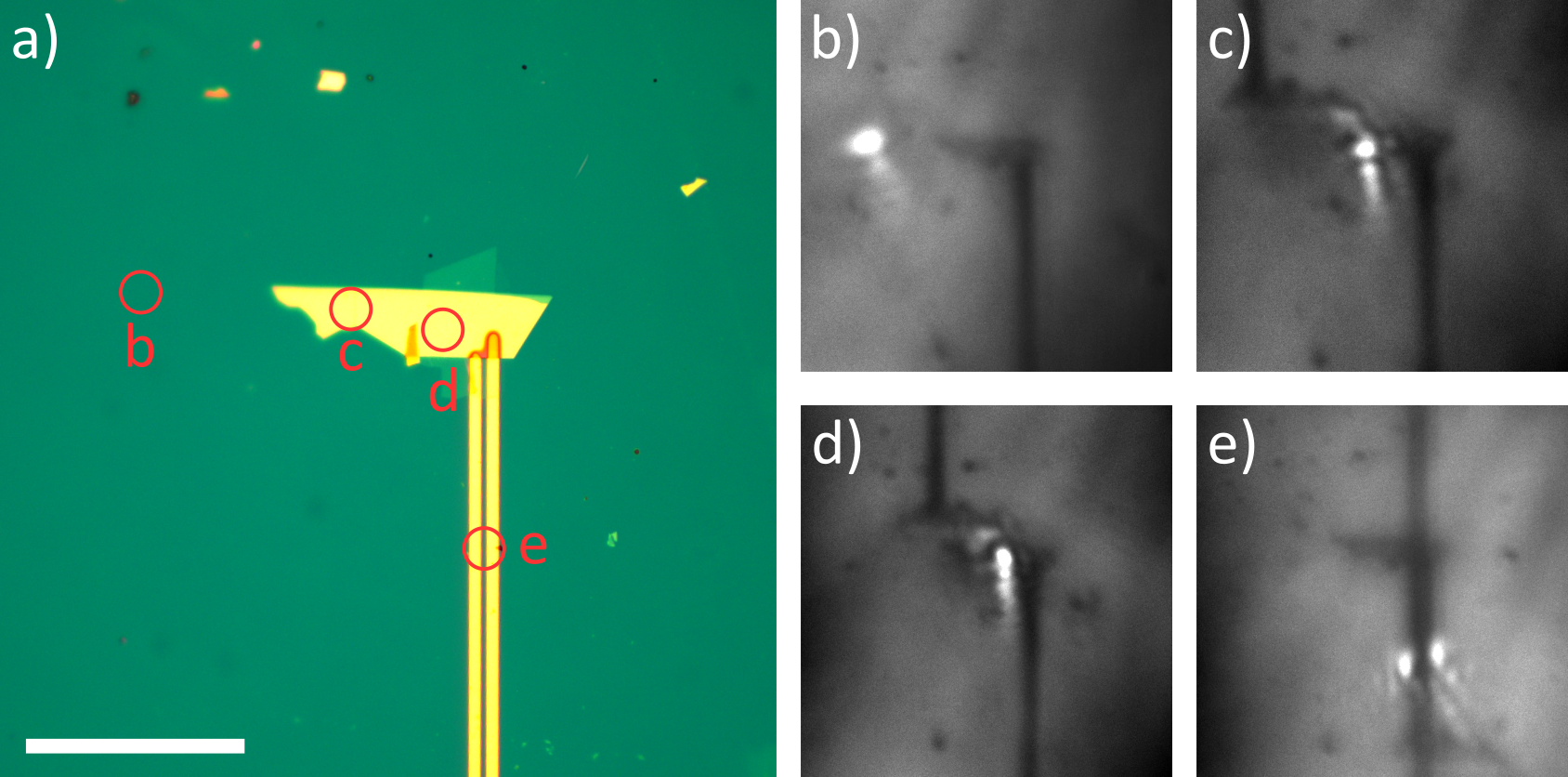}
    \caption{Optical images of a graphene-backgated TMD sample with gold leads. (a) Optical microscope image of the sample. Monolayer WSe$_2$ is hBN-encapsulated and connected to a graphite contact, with gold leads extending downward from the device. Red circles indicate the positions at which the cavity finesse measurements are made. (b-e) In situ, confocal-cavity images of the same sample region. A broadband LED source is resonantly coupled to cavity modes, and the cavity emission is imaged onto a camera.  The confocal supermode is positioned so that it passes through the: b) bare sapphire substrate; c) bare hBN region; d) TMD sample; and e) gold leads. The scale bar has a length of 50~$\mu$m.}
    \label{fig:sample_with_leads}
\end{figure}}

We measured the confocal cavity finesse using a supermode of waist $\approx 5~\mu\mathrm{m}$ that is positioned at different locations on the device.  A spatial light modulator is used to optimize the mode matching into the cavity. For the bare sapphire substrate, we find $\mathcal{F}\approx181$, while for the bare hBN region the finesse is $\mathcal{F}\approx120$; see Figs.~\ref{fig:sample_with_leads}b and c, respectively. For the TMD device region---Fig.~\ref{fig:sample_with_leads}d---the graphene backgate introduces additional optical absorption of approximately $2\%$, further reducing the finesse to $\mathcal{F}\approx80$.  Interestingly, when the mode is positioned at the midpoint between the gold leads but away from the graphene, see Fig.~\ref{fig:sample_with_leads}e, we still observe a finesse of ${\sim}100$. The confocal supermode self-consistently distorts around the metal leads to transmit through the gap and surrounding open areas. These measurements demonstrate that the cavity can accommodate electrically contacted 2D materials at the cavity midplane while retaining sufficient finesse for cavity-enhanced optical experiments.

\section{\label{sec:cryo}Cryogenics and Vacuum System}
We now describe the cryogenic and vacuum system, including how we vibrationally isolate the cryocooler.  Information about the radiation shields and thermal braids is also included.

\subsection{\label{sec:cryocooler}Closed-Cycle Cryocooler and
            Ultralow-Vibration Interface}

\noindent 
Cooling is provided by a two-stage pulse-tube closed-cycle cryocooler whose compressor is housed in an adjacent room to reduce acoustic noise at the experiment. The cold head is mounted on a ceiling rack and coupled to the chamber via flexible rubber bellows, attenuating the direct transmission of compressor vibrations.

Vibration transmission is the central challenge one faces when integrating a pulse-tube cryocooler with a high-finesse optical cavity. Without mitigation, the $\sim$1~Hz mechanical drive of the cold head and its harmonics would overwhelm one's ability to control the cavity length. We address this challenge using an ultralow-vibration interface (ULVI) in which heat is extracted from the cold finger by the cold head using a helium exchange-gas gap bridged between them by a copper-fin heat exchanger~\cite{Naides2013tug,Taylor2021asq}.
Because there is no rigid mechanical contact across this interface, only liquid and gaseous He cryogen, vibrations are effectively decoupled from the sample and cavity.  The dominant residual coupling is acoustic transmission through the exchange gas. The second-stage cold finger, reaching a base temperature of $\sim$4~K, is connected to the copper sample stick via flexible oxygen-free high-conductivity copper braids that conduct heat.  They remain sufficiently compliant to inhibit the transmission of the residual vibrations.

To maintain vacuum, we cannot employ a turbopump due to its vibrations.  Rather, after the initial pump down, the turbopump is shut off and an ion pump sustains UHV operation. The base pressure is $10^{-8}$~Torr at room temperature and $10^{-10}$~Torr under cryogenic conditions, with a pump-down time of approximately one day from ambient pressure.  The sample cooldown ramp can be performed concurrently with the latter portion of the vacuum pump down, for a roughly 24-hr time-to-base temperature and pressure.

\subsection{\label{sec:shield}Radiation Shields}

Our system relies on two sources of cooling for the heat shields.  A liquid nitrogen (LN$_2$) flow cryostat cools a copper radiation shield that surrounds the sample stage and cavity. This suppresses the blackbody heat load from room-temperature surfaces. It is shown in orange in Fig.~\ref{fig:apparatus_overview}c. This LN$_2$ heat shield cools to 110~K and is electro-polished to minimize thermal emissivity.

The second cooling source comes from the radiation shield of the cold finger extending from the pulse-tube cryostat.  This is thermally anchored to the mechanical housing for the cavity. The cavity mount and the mirrors that are coupled to it reach approximately 100~K; thus, the mirrors themselves serve as part of the 4$\pi$ heat shield for the intracavity sample. This heat shield also surrounds the cold finger from the pulse-tube cryostat, connecting to the cylindrical heat shield for this cold finger from the cryostat itself. This forms a continuous heat shield, as shown in yellow in Fig.~\ref{fig:apparatus_overview}b. Anchoring the cavity mount to this shield, rather than to a room-temperature structure, reduces the thermal gradient across the cavity mirror substrates and minimizes differential thermal expansion between them. The Macor posts that are used in the cavity housing and as spacers between the nanopositioners are thermally insulating and stiff; see the light gray shapes in Fig.~\ref{fig:cavity mount}a. 

Optical access through the shield is provided by antireflective sapphire windows that are heat sunk to the shield.  Sapphire is chosen for its high transmission across the visible and near-infrared, low room-temperature blackbody radiation transmission, and high thermal conductivity at cryogenic temperatures. Separate windows on the shield faces provide optical access to the cavity mirrors as well as access to two directions transverse to the cavity axis for pump beams.

 \subsection{Thermal Braids}

The thermal path from the ULVI cold finger to the sample is bridged by oxygen-free, high-conductivity copper braids. Braid flexibility is essential: A rigid link would reintroduce vibrations that the ULVI works to eliminate, while too little thermal conductance would limit the base temperature. We cool the sample stick by directly connecting a braid from the cold finger to the rotation nanopositioner, the nearest component to the sample stick.

The sample positioning stage is moved by piezoelectric slip-stick actuators.  These are sensitive to strain and the braids must be sufficiently compliant that their restoring force does not exceed the actuator's slip force. Otherwise, the range of the positioners will be limited. Braid cross-section, length, and routing are chosen to simultaneously satisfy the thermal, vibration isolation, and compliance requirements.

A resistive heater on the cold finger allows the sample temperature to be raised above base temperature. The sample temperature can be regulated from base temperature to 300~K.

\subsection{\label{sec:cooling}Sample Cooling}

The temperatures of the cryogenic assembly are monitored by seven calibrated resistance thermometers mounted at the ULVI cold finger, the cold finger radiation shield, the mirror housing above the cavity, the housing below the cavity, the LN$_2$ stem, the LN$_2$ radiation shield, and the sample stick. Together, these sensors map the temperature gradient from the cold head to the sample. 

With LN$_2$ flowing, the LN$_2$ radiation shield reaches a base temperature of approximately 110~K, suppressing the blackbody load on the inner assembly. The sample mount reaches a base temperature of 10.4~K, limited by the thermal resistance of the copper braids. Cooling takes roughly 10~hrs from room temperature if the LN$_2$ cold finger is already chilled. The one-day sample-exchange-and-pump-down time listed in Table~\ref{specstable} encompasses the LN2$_2$ pre-cooling and the 10-hr sample cool-down, with the cool-down running concurrently with the back end of the vacuum pump-down.

\section{\label{sec:detection}Imaging and Detection}

\subsection{\label{sec:insitu_imaging}In Situ Sample Imaging}

Because the cavity is multimode, it acts as a lens system, enabling intracavity materials to be imaged from cavity emission. We provide an example image of an intracavity TMD sample (hBN-encapsulated monolayer WSe$_2$) in Fig.~\ref{fig:sample_imaging}. For imaging (as opposed to cavity QED experiments) we use broad-spectrum incoherent light to reduce interference fringes.  Light emitted from the lower cavity mirror is collected with a lens and routed to a detection assembly outside the vacuum chamber. The emission can be diverted to any one of four channels: A CCD camera for in situ imaging; a bandpass-filtered, 780-nm photodiode for PDH locking; a broader 710-760-nm photodiode for detecting transmission of the longitudinal cavity pump at any relevant wavelength, and a single-photon counter for sensitive cavity emission measurements. 

To enable high numerical-aperture (NA) mode-matching to the cavity's supermode and for emission collection, two in-vacuum lenses are mounted above and below the cavity along its longitudinal axis.  The input lens focuses the in-coupling beam to a cavity supermode waist at the cavity midplane.  With an NA of $\sim$0.35, we have demonstrated cavity coupling down to a supermode waist of $\sim$5~$\mu$m in a test cavity using a spatial light modulator for mode matching. However,  the measurements presented above couple into a wider supermode of 9.5~$\mu$m.  In the future, we can also use a spatial light modulator to optimize the coupling into the cavity within the vacuum chamber and possibly achieve a 5-$\mu$m waist as well.  The output lens collimates the diverging emission beam before it exits the vacuum chamber, ensuring that the beam is not clipped by the heat shield or chamber windows.

The imaging channel provides a direct view of the sample through the cavity axis with a spatial resolution of  4~$\mu$m and a field of view of 500~$\mu$m. To maximize field intensity of the supermode at the sample, we set the sample position to be in the cavity midplane and centered about the cavity axis. Figure~\ref{fig:sample_imaging} shows representative images, including a wide field-of-view image containing the sample and surrounding features.

\section{\label{sec:char} Vibration characterization}

We now characterize the apparatus's vibrations, including measurements of cavity-length fluctuations and sample vibration with respect to the cavity mode.

\subsection{\label{sec:cav_vibs}Cavity Vibrations}

\begin{figure}
    \includegraphics[width=\columnwidth]{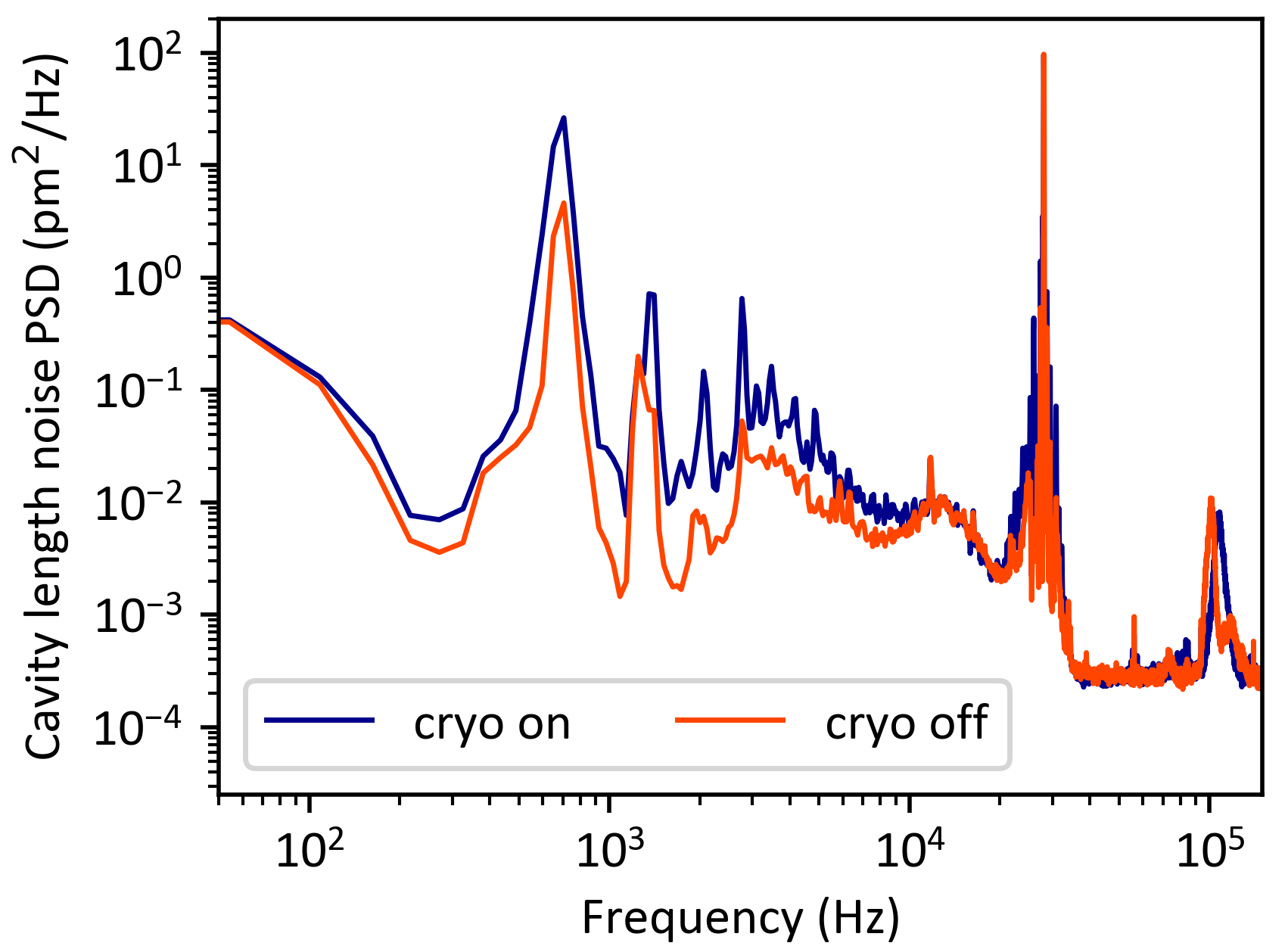}
    \caption{ Power spectral density (PSD) of cavity length noise with the cryocooler off [red, RMS vibrations $= 97(9)$~pm]
    and on [blue, RMS vibrations $=110(10)$~pm].  }
    \label{fig:cavity_vibrations}
\end{figure}

We characterize the cavity length stability by using the locked cavity itself as a displacement sensor.
With the PDH servo engaged, residual length fluctuations that the servo does not fully suppress appear as error signal fluctuations that we express as root-mean-square voltages $V_\text{RMS}$ over some specified bandwidth (30~Hz to $10^5$~Hz, as discussed below). We monitor this and convert it to a cavity length displacement $\delta L$ using the relation
\begin{equation}\label{deltaL}
    \delta L = \frac{V_\text{RMS}}{dV/d\nu} \cdot \frac{L_\text{cavity}}{q \cdot \nu_\text{FSR}},
\end{equation}
where $dV/d\nu$ is the open-loop error signal slope. $d\nu$ is calibrated by measuring the time between sidebands of known frequency written onto the cavity probe beam as the cavity length is swept in time. Therefore, the first factor $V_\text{RMS}/(dV/d\nu)$ is the RMS frequency deviation of residual vibrations of the locked cavity. Without a sample, this RMS deviation is less than 4.25~MHz, well within the cavity's 64-MHz linewidth. With a sample in the cavity, it is roughly 7~MHz.

In the second factor, $L_\text{cavity}$ is the length of the cavity and is approximately 10~mm. $\nu_\text{FSR} = c/(2L_\text{cavity})$ is the free spectral range, with $c$ the speed of light. Finally, $q = \nu_\text{laser}/\nu_\text{FSR}$ is the longitudinal mode number, i.e., the number of half-wavelengths of the laser of frequency $\nu_\text{laser}$ that span a standing wave of the cavity. The second factor in Eq.~\eqref{deltaL} is thus ${dL}/{d\nu}$. 

In Fig.~\ref{fig:cavity_vibrations}, we quantify the noise power spectral density (PSD) of these cavity length fluctuations.  Individual experimental measurements will last no longer than a few tens of ms, so we begin the frequency measurement at $f > 30$~Hz. We are interested in the steady-state response of the light-driven material, which evolves on timescales slower than ${\sim}10^5$~Hz. We therefore integrate the displacement noise PSD between these frequency scales as our measure of the cavity vibrations.

With the cryocooler off, the bare cavity exhibits a $\delta L = 97(9)$~pm RMS deviation, which we believe is dominated by lab acoustic and table vibrations. Turning on the cold head increases the RMS displacement by only 10 percent to $\delta L = 110(10)$~pm.  This demonstrates the efficacy of our cavity vibration isolation stack. Looking at the spectrum shown in Fig.~\ref{fig:cavity_vibrations}, we attribute the peak near 700~Hz to a mechanical resonance of the cavity mount, and the peaks past 10~kHz to the proportional-integral locking electronics.   For simplicity, both PSD plots are taken when the cavity length is tuned to realize a single-mode geometry. 

\subsection{Sample Vibrations}

\begin{figure}[!ht]
    \includegraphics[width=\columnwidth]{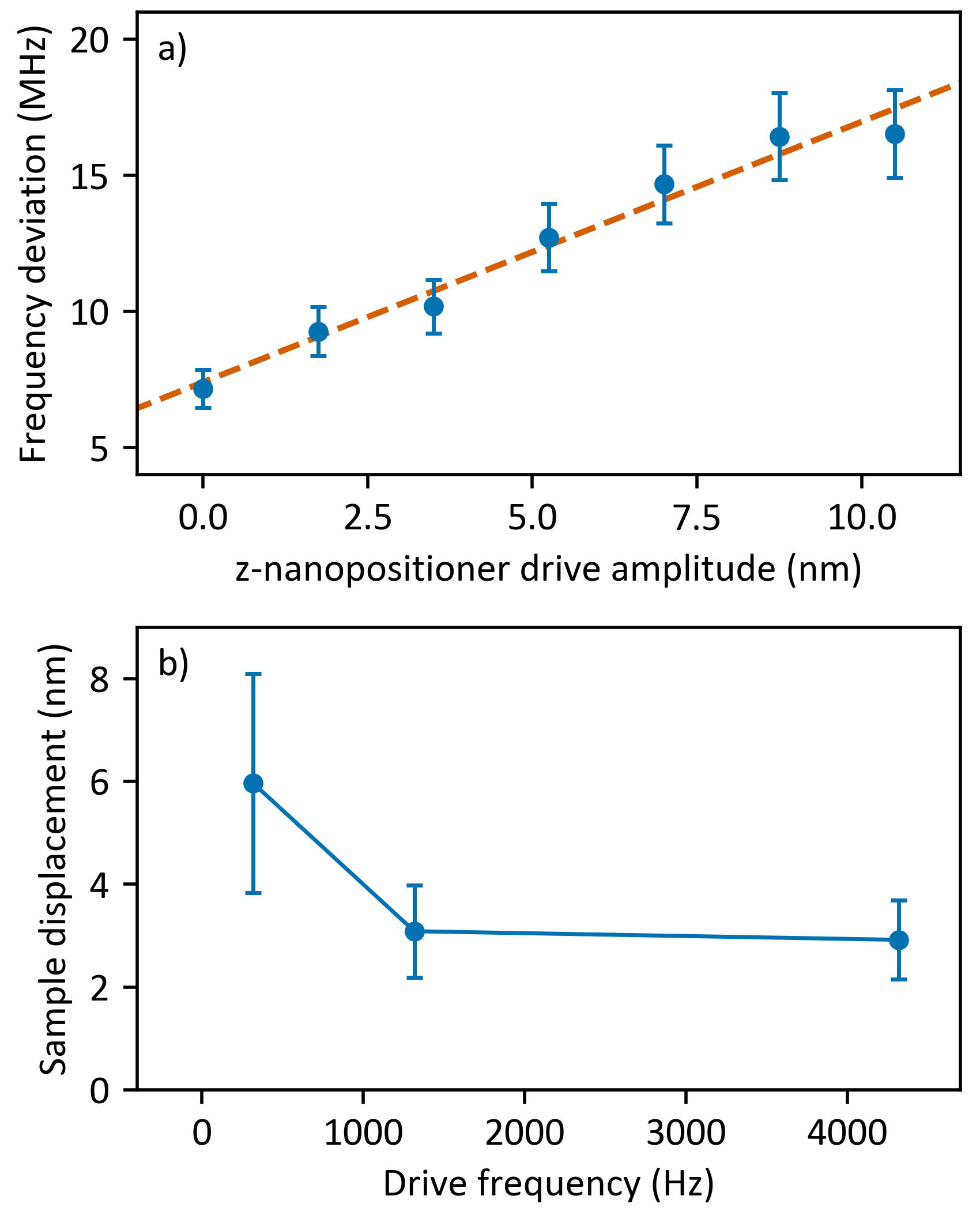}
    \caption{Calibration and results of the sample vibration measurement,
    all acquired with the cryocooler on and PDH lock engaged.
    (a)~RMS cavity frequency deviation versus driven substrate displacement at 1317~Hz. The linear fit (dashed) provides the displacement sensitivity $\alpha$ and zero-drive intercept $\beta$.
    (b)~Extracted intrinsic sample displacement at three drive frequencies, with uncertainties
    propagated from the linear-fit.}
    \label{fig:sample_vibrations}
\end{figure}

For cavity QED experiments, it is important that the sample motion remain small compared to the wavelength ($\lambda/2 \approx 390$~nm) of the intracavity standing wave.  This ensures that the electric field amplitude at the sample is sufficiently stable during a measurement. We quantify the intrinsic vibration of the sample stick with respect to the cavity mode position using the locked cavity as a displacement sensor. 

The measurement proceeds as follows. A piezo actuator in the $z$ axis stage beneath the sample stick is driven sinusoidally, displacing the sample along the cavity axis by a known amplitude $x = \gamma V_\text{drive}$, where $\gamma = 40.5\pm2.7$~nm/V. This voltage-to-displacement conversion factor is measured by scanning the piezo voltage and measuring the separation in voltage between transmission peaks separated by one FSR.

Figure~\ref{fig:sample_vibrations}a presents the RMS frequency deviation $\delta\nu$ of the cavity as a function of sample drive amplitude.  The response is approximately linear, which we parameterize as $\delta\nu = \alpha x + \beta$. The slope $\alpha$ is the cavity's displacement sensitivity, while the intercept $\beta$ accounts for residual noise. The contribution of the sample's motion to noise on the error signal is obtained by subtracting the independently measured bare-cavity noise floor $\delta\nu_\text{bare}\approx$ 4.25~MHz from $\beta$:
\begin{equation}
    x_\text{sample} = \frac{\beta - \delta\nu_\text{bare}}{\alpha}.
\end{equation}
Figure~\ref{fig:sample_vibrations}b presents these vibration measurements for three drive frequencies spanning 317--4317~Hz, all acquired with the cryocooler on and the PDH lock engaged. The extracted sample displacement remains at or below 10~nm RMS at all tested frequencies, well within our experimental requirement. We note that $\alpha$ (the displacement sensitivity) is not constant versus frequency. This is because the finite PDH servo bandwidth (around one kHz) changes the effective response of the cavity length to changes in  dispersive shift from substrate and sample movement within the standing wave. Uncertainties on $x_\text{sample}$ are obtained from the linear fit.


\section{\label{sec:conclusion}Conclusion}

We have developed a cryogenic, ultrahigh-vacuum platform for coupling 2D materials to a multimode optical cavity. The apparatus integrates a length-tunable confocal Fabry--P\'{e}rot cavity with in situ optical access, electrical transport capability, and cryogenic nanopositioning. This system enables simultaneous optical and electronic measurements.  Moreover, cavity length and sample-cavity relative vibrations are minimized. Operation near the confocal condition produces localized supermodes, enhancing the light–matter interaction strength. Future work will measure the enhancement factor through polariton spectroscopy.  With the cryocooler on, cavity length noise is no larger than $\sim$110~pm RMS and sample motion with respect to the cavity standing-wave mode structure is less than $10$~nm RMS. This ensures stable sample-cavity coupling, even at cryogenic temperatures down to $10.4$~K. The open cavity geometry enables transverse pumping while maintaining tight mode confinement.

\begin{acknowledgments}
We thank Jonathan Keeling and Qitong Li for helpful discussions.  The Stanford work is supported by the Gordon and Betty Moore Foundation (grant number GBMF10693).  The work at U.~Washington is supported by DoE-BES under the award DE-SC0018171.
\end{acknowledgments}


\begin{thebibliography}{46}%
\makeatletter
\providecommand \@ifxundefined [1]{%
 \@ifx{#1\undefined}
}%
\providecommand \@ifnum [1]{%
 \ifnum #1\expandafter \@firstoftwo
 \else \expandafter \@secondoftwo
 \fi
}%
\providecommand \@ifx [1]{%
 \ifx #1\expandafter \@firstoftwo
 \else \expandafter \@secondoftwo
 \fi
}%
\providecommand \natexlab [1]{#1}%
\providecommand \enquote  [1]{``#1''}%
\providecommand \bibnamefont  [1]{#1}%
\providecommand \bibfnamefont [1]{#1}%
\providecommand \citenamefont [1]{#1}%
\providecommand \href@noop [0]{\@secondoftwo}%
\providecommand \href [0]{\begingroup \@sanitize@url \@href}%
\providecommand \@href[1]{\@@startlink{#1}\@@href}%
\providecommand \@@href[1]{\endgroup#1\@@endlink}%
\providecommand \@sanitize@url [0]{\catcode `\\12\catcode `\$12\catcode `\&12\catcode `\#12\catcode `\^12\catcode `\_12\catcode `\%12\relax}%
\providecommand \@@startlink[1]{}%
\providecommand \@@endlink[0]{}%
\providecommand \url  [0]{\begingroup\@sanitize@url \@url }%
\providecommand \@url [1]{\endgroup\@href {#1}{\urlprefix }}%
\providecommand \urlprefix  [0]{URL }%
\providecommand \Eprint [0]{\href }%
\providecommand \doibase [0]{http://dx.doi.org/}%
\providecommand \selectlanguage [0]{\@gobble}%
\providecommand \bibinfo  [0]{\@secondoftwo}%
\providecommand \bibfield  [0]{\@secondoftwo}%
\providecommand \translation [1]{[#1]}%
\providecommand \BibitemOpen [0]{}%
\providecommand \bibitemStop [0]{}%
\providecommand \bibitemNoStop [0]{.\EOS\space}%
\providecommand \EOS [0]{\spacefactor3000\relax}%
\providecommand \BibitemShut  [1]{\csname bibitem#1\endcsname}%
\let\auto@bib@innerbib\@empty
\bibitem [{\citenamefont {Geim}\ and\ \citenamefont {Grigorieva}(2013)}]{Geim2013}%
  \BibitemOpen
  \bibfield  {author} {\bibinfo {author} {\bibfnamefont {A.~K.}\ \bibnamefont {Geim}}\ and\ \bibinfo {author} {\bibfnamefont {I.~V.}\ \bibnamefont {Grigorieva}},\ }\bibfield  {title} {\enquote {\bibinfo {title} {{Van der Waals heterostructures}},}\ }\href {\doibase 10.1038/nature12385} {\bibfield  {journal} {\bibinfo  {journal} {Nature}\ }\textbf {\bibinfo {volume} {499}},\ \bibinfo {pages} {419--425} (\bibinfo {year} {2013})}\BibitemShut {NoStop}%
\bibitem [{\citenamefont {Cao}\ \emph {et~al.}(2018)\citenamefont {Cao}, \citenamefont {Fatemi}, \citenamefont {Fang}, \citenamefont {Watanabe}, \citenamefont {Taniguchi}, \citenamefont {Kaxiras},\ and\ \citenamefont {Jarillo-Herrero}}]{Cao2018}%
  \BibitemOpen
  \bibfield  {author} {\bibinfo {author} {\bibfnamefont {Y.}~\bibnamefont {Cao}}, \bibinfo {author} {\bibfnamefont {V.}~\bibnamefont {Fatemi}}, \bibinfo {author} {\bibfnamefont {S.}~\bibnamefont {Fang}}, \bibinfo {author} {\bibfnamefont {K.}~\bibnamefont {Watanabe}}, \bibinfo {author} {\bibfnamefont {T.}~\bibnamefont {Taniguchi}}, \bibinfo {author} {\bibfnamefont {E.}~\bibnamefont {Kaxiras}}, \ and\ \bibinfo {author} {\bibfnamefont {P.}~\bibnamefont {Jarillo-Herrero}},\ }\bibfield  {title} {\enquote {\bibinfo {title} {{Unconventional superconductivity in magic-angle graphene superlattices}},}\ }\href {\doibase 10.1038/nature26160} {\bibfield  {journal} {\bibinfo  {journal} {Nature}\ }\textbf {\bibinfo {volume} {556}},\ \bibinfo {pages} {43--50} (\bibinfo {year} {2018})}\BibitemShut {NoStop}%
\bibitem [{\citenamefont {Xi}\ \emph {et~al.}(2016)\citenamefont {Xi}, \citenamefont {Wang}, \citenamefont {Zhao}, \citenamefont {Park}, \citenamefont {Law}, \citenamefont {Berger}, \citenamefont {Forr\'{o}}, \citenamefont {Shan},\ and\ \citenamefont {Mak}}]{Xi2016}%
  \BibitemOpen
  \bibfield  {author} {\bibinfo {author} {\bibfnamefont {X.}~\bibnamefont {Xi}}, \bibinfo {author} {\bibfnamefont {Z.}~\bibnamefont {Wang}}, \bibinfo {author} {\bibfnamefont {W.}~\bibnamefont {Zhao}}, \bibinfo {author} {\bibfnamefont {J.-H.}\ \bibnamefont {Park}}, \bibinfo {author} {\bibfnamefont {K.~T.}\ \bibnamefont {Law}}, \bibinfo {author} {\bibfnamefont {H.}~\bibnamefont {Berger}}, \bibinfo {author} {\bibfnamefont {L.}~\bibnamefont {Forr\'{o}}}, \bibinfo {author} {\bibfnamefont {J.}~\bibnamefont {Shan}}, \ and\ \bibinfo {author} {\bibfnamefont {K.~F.}\ \bibnamefont {Mak}},\ }\bibfield  {title} {\enquote {\bibinfo {title} {{Ising pairing in superconducting NbSe$_2$ atomic layers}},}\ }\href {\doibase 10.1038/nphys3538} {\bibfield  {journal} {\bibinfo  {journal} {Nat. Phys.}\ }\textbf {\bibinfo {volume} {12}},\ \bibinfo {pages} {139--143} (\bibinfo {year} {2016})}\BibitemShut {NoStop}%
\bibitem [{\citenamefont {Lu}\ \emph {et~al.}(2019)\citenamefont {Lu}, \citenamefont {Stepanov}, \citenamefont {Yang}, \citenamefont {Xie}, \citenamefont {Aamir}, \citenamefont {Das}, \citenamefont {Urgell}, \citenamefont {Watanabe}, \citenamefont {Taniguchi}, \citenamefont {Zhang}, \citenamefont {Bachtold}, \citenamefont {MacDonald},\ and\ \citenamefont {Efetov}}]{Lu2019}%
  \BibitemOpen
  \bibfield  {author} {\bibinfo {author} {\bibfnamefont {X.}~\bibnamefont {Lu}}, \bibinfo {author} {\bibfnamefont {P.}~\bibnamefont {Stepanov}}, \bibinfo {author} {\bibfnamefont {W.}~\bibnamefont {Yang}}, \bibinfo {author} {\bibfnamefont {M.}~\bibnamefont {Xie}}, \bibinfo {author} {\bibfnamefont {M.~A.}\ \bibnamefont {Aamir}}, \bibinfo {author} {\bibfnamefont {I.}~\bibnamefont {Das}}, \bibinfo {author} {\bibfnamefont {C.}~\bibnamefont {Urgell}}, \bibinfo {author} {\bibfnamefont {K.}~\bibnamefont {Watanabe}}, \bibinfo {author} {\bibfnamefont {T.}~\bibnamefont {Taniguchi}}, \bibinfo {author} {\bibfnamefont {G.}~\bibnamefont {Zhang}}, \bibinfo {author} {\bibfnamefont {A.}~\bibnamefont {Bachtold}}, \bibinfo {author} {\bibfnamefont {A.~H.}\ \bibnamefont {MacDonald}}, \ and\ \bibinfo {author} {\bibfnamefont {D.~K.}\ \bibnamefont {Efetov}},\ }\bibfield  {title} {\enquote {\bibinfo {title} {{Superconductors, orbital magnets and correlated states in magic-angle bilayer graphene}},}\ }\href {\doibase
  10.1038/s41586-019-1695-0} {\bibfield  {journal} {\bibinfo  {journal} {Nature}\ }\textbf {\bibinfo {volume} {574}},\ \bibinfo {pages} {653--657} (\bibinfo {year} {2019})}\BibitemShut {NoStop}%
\bibitem [{\citenamefont {Fei}\ \emph {et~al.}(2017)\citenamefont {Fei}, \citenamefont {Palomaki}, \citenamefont {Wu}, \citenamefont {Zhao}, \citenamefont {Cai}, \citenamefont {Sun}, \citenamefont {Nguyen}, \citenamefont {Finney}, \citenamefont {Xu},\ and\ \citenamefont {Cobden}}]{Fei2017}%
  \BibitemOpen
  \bibfield  {author} {\bibinfo {author} {\bibfnamefont {Z.}~\bibnamefont {Fei}}, \bibinfo {author} {\bibfnamefont {T.}~\bibnamefont {Palomaki}}, \bibinfo {author} {\bibfnamefont {S.}~\bibnamefont {Wu}}, \bibinfo {author} {\bibfnamefont {W.}~\bibnamefont {Zhao}}, \bibinfo {author} {\bibfnamefont {X.}~\bibnamefont {Cai}}, \bibinfo {author} {\bibfnamefont {B.}~\bibnamefont {Sun}}, \bibinfo {author} {\bibfnamefont {P.}~\bibnamefont {Nguyen}}, \bibinfo {author} {\bibfnamefont {J.}~\bibnamefont {Finney}}, \bibinfo {author} {\bibfnamefont {X.}~\bibnamefont {Xu}}, \ and\ \bibinfo {author} {\bibfnamefont {D.~H.}\ \bibnamefont {Cobden}},\ }\bibfield  {title} {\enquote {\bibinfo {title} {{Edge conduction in monolayer WTe$_2$}},}\ }\href {\doibase 10.1038/nphys4091} {\bibfield  {journal} {\bibinfo  {journal} {Nat. Phys.}\ }\textbf {\bibinfo {volume} {13}},\ \bibinfo {pages} {677--682} (\bibinfo {year} {2017})}\BibitemShut {NoStop}%
\bibitem [{\citenamefont {Sajadi}\ \emph {et~al.}(2018)\citenamefont {Sajadi}, \citenamefont {Palomaki}, \citenamefont {Fei}, \citenamefont {Zhao}, \citenamefont {Bement}, \citenamefont {Olsen}, \citenamefont {Luescher}, \citenamefont {Xu}, \citenamefont {Folk},\ and\ \citenamefont {Cobden}}]{Sajadi2018}%
  \BibitemOpen
  \bibfield  {author} {\bibinfo {author} {\bibfnamefont {E.}~\bibnamefont {Sajadi}}, \bibinfo {author} {\bibfnamefont {T.}~\bibnamefont {Palomaki}}, \bibinfo {author} {\bibfnamefont {Z.}~\bibnamefont {Fei}}, \bibinfo {author} {\bibfnamefont {W.}~\bibnamefont {Zhao}}, \bibinfo {author} {\bibfnamefont {P.}~\bibnamefont {Bement}}, \bibinfo {author} {\bibfnamefont {C.}~\bibnamefont {Olsen}}, \bibinfo {author} {\bibfnamefont {S.}~\bibnamefont {Luescher}}, \bibinfo {author} {\bibfnamefont {X.}~\bibnamefont {Xu}}, \bibinfo {author} {\bibfnamefont {J.~A.}\ \bibnamefont {Folk}}, \ and\ \bibinfo {author} {\bibfnamefont {D.~H.}\ \bibnamefont {Cobden}},\ }\bibfield  {title} {\enquote {\bibinfo {title} {{Gate-induced superconductivity in a monolayer topological insulator}},}\ }\href {\doibase 10.1126/science.aar4426} {\bibfield  {journal} {\bibinfo  {journal} {Science}\ }\textbf {\bibinfo {volume} {362}},\ \bibinfo {pages} {922--925} (\bibinfo {year} {2018})}\BibitemShut {NoStop}%
\bibitem [{\citenamefont {Sharpe}\ \emph {et~al.}(2019)\citenamefont {Sharpe}, \citenamefont {Fox}, \citenamefont {Barnard}, \citenamefont {Finney}, \citenamefont {Watanabe}, \citenamefont {Taniguchi}, \citenamefont {Kastner},\ and\ \citenamefont {Goldhaber-Gordon}}]{Sharpe2019}%
  \BibitemOpen
  \bibfield  {author} {\bibinfo {author} {\bibfnamefont {A.~L.}\ \bibnamefont {Sharpe}}, \bibinfo {author} {\bibfnamefont {E.~J.}\ \bibnamefont {Fox}}, \bibinfo {author} {\bibfnamefont {A.~W.}\ \bibnamefont {Barnard}}, \bibinfo {author} {\bibfnamefont {J.}~\bibnamefont {Finney}}, \bibinfo {author} {\bibfnamefont {K.}~\bibnamefont {Watanabe}}, \bibinfo {author} {\bibfnamefont {T.}~\bibnamefont {Taniguchi}}, \bibinfo {author} {\bibfnamefont {M.~A.}\ \bibnamefont {Kastner}}, \ and\ \bibinfo {author} {\bibfnamefont {D.}~\bibnamefont {Goldhaber-Gordon}},\ }\bibfield  {title} {\enquote {\bibinfo {title} {Emergent ferromagnetism near three-quarters filling in twisted bilayer graphene},}\ }\href {\doibase 10.1126/science.aaw3780} {\bibfield  {journal} {\bibinfo  {journal} {Science}\ }\textbf {\bibinfo {volume} {365}},\ \bibinfo {pages} {605--608} (\bibinfo {year} {2019})}\BibitemShut {NoStop}%
\bibitem [{\citenamefont {Huang}\ \emph {et~al.}(2017)\citenamefont {Huang}, \citenamefont {Clark}, \citenamefont {Navarro-Moratalla}, \citenamefont {Klein}, \citenamefont {Cheng}, \citenamefont {Seyler}, \citenamefont {Zhong}, \citenamefont {Schmidgall}, \citenamefont {McGuire}, \citenamefont {Cobden}, \citenamefont {Yao}, \citenamefont {Xiao}, \citenamefont {Jarillo-Herrero},\ and\ \citenamefont {Xu}}]{Huang2017}%
  \BibitemOpen
  \bibfield  {author} {\bibinfo {author} {\bibfnamefont {B.}~\bibnamefont {Huang}}, \bibinfo {author} {\bibfnamefont {G.}~\bibnamefont {Clark}}, \bibinfo {author} {\bibfnamefont {E.}~\bibnamefont {Navarro-Moratalla}}, \bibinfo {author} {\bibfnamefont {D.~R.}\ \bibnamefont {Klein}}, \bibinfo {author} {\bibfnamefont {R.}~\bibnamefont {Cheng}}, \bibinfo {author} {\bibfnamefont {K.~L.}\ \bibnamefont {Seyler}}, \bibinfo {author} {\bibfnamefont {D.}~\bibnamefont {Zhong}}, \bibinfo {author} {\bibfnamefont {E.}~\bibnamefont {Schmidgall}}, \bibinfo {author} {\bibfnamefont {M.~A.}\ \bibnamefont {McGuire}}, \bibinfo {author} {\bibfnamefont {D.~H.}\ \bibnamefont {Cobden}}, \bibinfo {author} {\bibfnamefont {W.}~\bibnamefont {Yao}}, \bibinfo {author} {\bibfnamefont {D.}~\bibnamefont {Xiao}}, \bibinfo {author} {\bibfnamefont {P.}~\bibnamefont {Jarillo-Herrero}}, \ and\ \bibinfo {author} {\bibfnamefont {X.}~\bibnamefont {Xu}},\ }\bibfield  {title} {\enquote {\bibinfo {title} {{Layer-dependent ferromagnetism in a van der
  Waals crystal down to the monolayer limit}},}\ }\href {\doibase 10.1038/nature22391} {\bibfield  {journal} {\bibinfo  {journal} {Nature}\ }\textbf {\bibinfo {volume} {546}},\ \bibinfo {pages} {270--273} (\bibinfo {year} {2017})}\BibitemShut {NoStop}%
\bibitem [{\citenamefont {Mak}\ and\ \citenamefont {Shan}(2022)}]{Mak2022smm}%
  \BibitemOpen
  \bibfield  {author} {\bibinfo {author} {\bibfnamefont {K.~F.}\ \bibnamefont {Mak}}\ and\ \bibinfo {author} {\bibfnamefont {J.}~\bibnamefont {Shan}},\ }\bibfield  {title} {\enquote {\bibinfo {title} {{Semiconductor moiré materials}},}\ }\href {\doibase 10.1038/s41565-022-01165-6} {\bibfield  {journal} {\bibinfo  {journal} {Nat. Nanotechnol.}\ }\textbf {\bibinfo {volume} {17}},\ \bibinfo {pages} {686--695} (\bibinfo {year} {2022})}\BibitemShut {NoStop}%
\bibitem [{\citenamefont {Fausti}\ \emph {et~al.}(2011)\citenamefont {Fausti}, \citenamefont {Tobey}, \citenamefont {Dean}, \citenamefont {Kaiser}, \citenamefont {Dienst}, \citenamefont {Hoffmann}, \citenamefont {Pyon}, \citenamefont {Takayama}, \citenamefont {Takagi},\ and\ \citenamefont {Cavalleri}}]{Fausti2011lsi}%
  \BibitemOpen
  \bibfield  {author} {\bibinfo {author} {\bibfnamefont {D.}~\bibnamefont {Fausti}}, \bibinfo {author} {\bibfnamefont {R.~I.}\ \bibnamefont {Tobey}}, \bibinfo {author} {\bibfnamefont {N.}~\bibnamefont {Dean}}, \bibinfo {author} {\bibfnamefont {S.}~\bibnamefont {Kaiser}}, \bibinfo {author} {\bibfnamefont {A.}~\bibnamefont {Dienst}}, \bibinfo {author} {\bibfnamefont {M.~C.}\ \bibnamefont {Hoffmann}}, \bibinfo {author} {\bibfnamefont {S.}~\bibnamefont {Pyon}}, \bibinfo {author} {\bibfnamefont {T.}~\bibnamefont {Takayama}}, \bibinfo {author} {\bibfnamefont {H.}~\bibnamefont {Takagi}}, \ and\ \bibinfo {author} {\bibfnamefont {A.}~\bibnamefont {Cavalleri}},\ }\bibfield  {title} {\enquote {\bibinfo {title} {{Light-Induced Superconductivity in a Stripe-Ordered Cuprate}},}\ }\href {\doibase 10.1126/science.1197294} {\bibfield  {journal} {\bibinfo  {journal} {Science}\ }\textbf {\bibinfo {volume} {331}},\ \bibinfo {pages} {189--191} (\bibinfo {year} {2011})}\BibitemShut {NoStop}%
\bibitem [{\citenamefont {Mankowsky}\ \emph {et~al.}(2014)\citenamefont {Mankowsky}, \citenamefont {Subedi}, \citenamefont {F\"orst}, \citenamefont {Mariager}, \citenamefont {Chollet}, \citenamefont {Lemke}, \citenamefont {Robinson}, \citenamefont {Glownia}, \citenamefont {Minitti}, \citenamefont {Frano}, \citenamefont {Fechner}, \citenamefont {Spaldin}, \citenamefont {Loew}, \citenamefont {Keimer}, \citenamefont {Georges},\ and\ \citenamefont {Cavalleri}}]{Mankowsky2014nld}%
  \BibitemOpen
  \bibfield  {author} {\bibinfo {author} {\bibfnamefont {R.}~\bibnamefont {Mankowsky}}, \bibinfo {author} {\bibfnamefont {A.}~\bibnamefont {Subedi}}, \bibinfo {author} {\bibfnamefont {M.}~\bibnamefont {F\"orst}}, \bibinfo {author} {\bibfnamefont {S.~O.}\ \bibnamefont {Mariager}}, \bibinfo {author} {\bibfnamefont {M.}~\bibnamefont {Chollet}}, \bibinfo {author} {\bibfnamefont {H.~T.}\ \bibnamefont {Lemke}}, \bibinfo {author} {\bibfnamefont {J.~S.}\ \bibnamefont {Robinson}}, \bibinfo {author} {\bibfnamefont {J.~M.}\ \bibnamefont {Glownia}}, \bibinfo {author} {\bibfnamefont {M.~P.}\ \bibnamefont {Minitti}}, \bibinfo {author} {\bibfnamefont {A.}~\bibnamefont {Frano}}, \bibinfo {author} {\bibfnamefont {M.}~\bibnamefont {Fechner}}, \bibinfo {author} {\bibfnamefont {N.~A.}\ \bibnamefont {Spaldin}}, \bibinfo {author} {\bibfnamefont {T.}~\bibnamefont {Loew}}, \bibinfo {author} {\bibfnamefont {B.}~\bibnamefont {Keimer}}, \bibinfo {author} {\bibfnamefont {A.}~\bibnamefont {Georges}}, \ and\ \bibinfo {author}
  {\bibfnamefont {A.}~\bibnamefont {Cavalleri}},\ }\bibfield  {title} {\enquote {\bibinfo {title} {{Nonlinear lattice dynamics as a basis for enhanced superconductivity in YBa2Cu3O6.5}},}\ }\href {\doibase 10.1038/nature13875} {\bibfield  {journal} {\bibinfo  {journal} {Nature}\ }\textbf {\bibinfo {volume} {516}},\ \bibinfo {pages} {71--73} (\bibinfo {year} {2014})}\BibitemShut {NoStop}%
\bibitem [{\citenamefont {Mitrano}\ \emph {et~al.}(2016)\citenamefont {Mitrano}, \citenamefont {Cantaluppi}, \citenamefont {Nicoletti}, \citenamefont {Kaiser}, \citenamefont {Perucchi}, \citenamefont {Lupi}, \citenamefont {Di~Pietro}, \citenamefont {Pontiroli}, \citenamefont {Ricc\`o}, \citenamefont {Clark}, \citenamefont {Jaksch},\ and\ \citenamefont {Cavalleri}}]{Mitrano2016pls}%
  \BibitemOpen
  \bibfield  {author} {\bibinfo {author} {\bibfnamefont {M.}~\bibnamefont {Mitrano}}, \bibinfo {author} {\bibfnamefont {A.}~\bibnamefont {Cantaluppi}}, \bibinfo {author} {\bibfnamefont {D.}~\bibnamefont {Nicoletti}}, \bibinfo {author} {\bibfnamefont {S.}~\bibnamefont {Kaiser}}, \bibinfo {author} {\bibfnamefont {A.}~\bibnamefont {Perucchi}}, \bibinfo {author} {\bibfnamefont {S.}~\bibnamefont {Lupi}}, \bibinfo {author} {\bibfnamefont {P.}~\bibnamefont {Di~Pietro}}, \bibinfo {author} {\bibfnamefont {D.}~\bibnamefont {Pontiroli}}, \bibinfo {author} {\bibfnamefont {M.}~\bibnamefont {Ricc\`o}}, \bibinfo {author} {\bibfnamefont {S.~R.}\ \bibnamefont {Clark}}, \bibinfo {author} {\bibfnamefont {D.}~\bibnamefont {Jaksch}}, \ and\ \bibinfo {author} {\bibfnamefont {A.}~\bibnamefont {Cavalleri}},\ }\bibfield  {title} {\enquote {\bibinfo {title} {{Possible light-induced superconductivity in K3C60 at high temperature}},}\ }\href {\doibase 10.1038/nature16522} {\bibfield  {journal} {\bibinfo  {journal} {Nature}\ }\textbf
  {\bibinfo {volume} {530}},\ \bibinfo {pages} {461--464} (\bibinfo {year} {2016})}\BibitemShut {NoStop}%
\bibitem [{\citenamefont {Cavalleri}(2017)}]{Cavalleri2017ps}%
  \BibitemOpen
  \bibfield  {author} {\bibinfo {author} {\bibfnamefont {A.}~\bibnamefont {Cavalleri}},\ }\bibfield  {title} {\enquote {\bibinfo {title} {{Photo-induced superconductivity}},}\ }\href {\doibase 10.1080/00107514.2017.1406623} {\bibfield  {journal} {\bibinfo  {journal} {Contemp. Phys.}\ }\textbf {\bibinfo {volume} {59}},\ \bibinfo {pages} {31--46} (\bibinfo {year} {2017})}\BibitemShut {NoStop}%
\bibitem [{\citenamefont {Disa}, \citenamefont {Nova},\ and\ \citenamefont {Cavalleri}(2021)}]{Disa2021ecs}%
  \BibitemOpen
  \bibfield  {author} {\bibinfo {author} {\bibfnamefont {A.~S.}\ \bibnamefont {Disa}}, \bibinfo {author} {\bibfnamefont {T.~F.}\ \bibnamefont {Nova}}, \ and\ \bibinfo {author} {\bibfnamefont {A.}~\bibnamefont {Cavalleri}},\ }\bibfield  {title} {\enquote {\bibinfo {title} {{Engineering crystal structures with light}},}\ }\href {\doibase 10.1038/s41567-021-01366-1} {\bibfield  {journal} {\bibinfo  {journal} {Nat. Phys.}\ }\textbf {\bibinfo {volume} {17}},\ \bibinfo {pages} {1087--1092} (\bibinfo {year} {2021})}\BibitemShut {NoStop}%
\bibitem [{\citenamefont {Bourzutschky}, \citenamefont {Lev},\ and\ \citenamefont {Keeling}(2024)}]{Bourzutschky2024rci}%
  \BibitemOpen
  \bibfield  {author} {\bibinfo {author} {\bibfnamefont {A.~N.}\ \bibnamefont {Bourzutschky}}, \bibinfo {author} {\bibfnamefont {B.~L.}\ \bibnamefont {Lev}}, \ and\ \bibinfo {author} {\bibfnamefont {J.}~\bibnamefont {Keeling}},\ }\bibfield  {title} {\enquote {\bibinfo {title} {{Raman-phonon-polariton condensation in a transversely pumped cavity}},}\ }\href {\doibase 10.1038/s41535-024-00693-9} {\bibfield  {journal} {\bibinfo  {journal} {npj Quantum Mater.}\ }\textbf {\bibinfo {volume} {9}},\ \bibinfo {pages} {81} (\bibinfo {year} {2024})}\BibitemShut {NoStop}%
\bibitem [{\citenamefont {Chernikov}\ \emph {et~al.}(2014)\citenamefont {Chernikov}, \citenamefont {Berkelbach}, \citenamefont {Hill}, \citenamefont {Rigosi}, \citenamefont {Li}, \citenamefont {Aslan}, \citenamefont {Reichman}, \citenamefont {Hybertsen},\ and\ \citenamefont {Heinz}}]{Chernikov2014}%
  \BibitemOpen
  \bibfield  {author} {\bibinfo {author} {\bibfnamefont {A.}~\bibnamefont {Chernikov}}, \bibinfo {author} {\bibfnamefont {T.~C.}\ \bibnamefont {Berkelbach}}, \bibinfo {author} {\bibfnamefont {H.~M.}\ \bibnamefont {Hill}}, \bibinfo {author} {\bibfnamefont {A.}~\bibnamefont {Rigosi}}, \bibinfo {author} {\bibfnamefont {Y.}~\bibnamefont {Li}}, \bibinfo {author} {\bibfnamefont {B.}~\bibnamefont {Aslan}}, \bibinfo {author} {\bibfnamefont {D.~R.}\ \bibnamefont {Reichman}}, \bibinfo {author} {\bibfnamefont {M.~S.}\ \bibnamefont {Hybertsen}}, \ and\ \bibinfo {author} {\bibfnamefont {T.~F.}\ \bibnamefont {Heinz}},\ }\bibfield  {title} {\enquote {\bibinfo {title} {{Exciton binding energy and nonhydrogenic Rydberg series in monolayer WS$_2$}},}\ }\href {\doibase 10.1103/PhysRevLett.113.076802} {\bibfield  {journal} {\bibinfo  {journal} {Phys. Rev. Lett.}\ }\textbf {\bibinfo {volume} {113}},\ \bibinfo {pages} {076802} (\bibinfo {year} {2014})}\BibitemShut {NoStop}%
\bibitem [{\citenamefont {Cadiz}\ \emph {et~al.}(2017)\citenamefont {Cadiz}, \citenamefont {Courtade}, \citenamefont {Robert}, \citenamefont {Wang}, \citenamefont {Shen}, \citenamefont {Cai}, \citenamefont {Taniguchi}, \citenamefont {Watanabe}, \citenamefont {Carrere}, \citenamefont {Lagarde}, \citenamefont {Manca}, \citenamefont {Amand}, \citenamefont {Renucci}, \citenamefont {Tongay}, \citenamefont {Marie},\ and\ \citenamefont {Urbaszek}}]{Cadiz2017}%
  \BibitemOpen
  \bibfield  {author} {\bibinfo {author} {\bibfnamefont {F.}~\bibnamefont {Cadiz}}, \bibinfo {author} {\bibfnamefont {E.}~\bibnamefont {Courtade}}, \bibinfo {author} {\bibfnamefont {C.}~\bibnamefont {Robert}}, \bibinfo {author} {\bibfnamefont {G.}~\bibnamefont {Wang}}, \bibinfo {author} {\bibfnamefont {Y.}~\bibnamefont {Shen}}, \bibinfo {author} {\bibfnamefont {H.}~\bibnamefont {Cai}}, \bibinfo {author} {\bibfnamefont {T.}~\bibnamefont {Taniguchi}}, \bibinfo {author} {\bibfnamefont {K.}~\bibnamefont {Watanabe}}, \bibinfo {author} {\bibfnamefont {H.}~\bibnamefont {Carrere}}, \bibinfo {author} {\bibfnamefont {D.}~\bibnamefont {Lagarde}}, \bibinfo {author} {\bibfnamefont {M.}~\bibnamefont {Manca}}, \bibinfo {author} {\bibfnamefont {T.}~\bibnamefont {Amand}}, \bibinfo {author} {\bibfnamefont {P.}~\bibnamefont {Renucci}}, \bibinfo {author} {\bibfnamefont {S.}~\bibnamefont {Tongay}}, \bibinfo {author} {\bibfnamefont {X.}~\bibnamefont {Marie}}, \ and\ \bibinfo {author} {\bibfnamefont {B.}~\bibnamefont {Urbaszek}},\
  }\bibfield  {title} {\enquote {\bibinfo {title} {{Excitonic linewidth approaching the homogeneous limit in MoS$_2$-based van der Waals heterostructures}},}\ }\href {\doibase 10.1103/PhysRevX.7.021026} {\bibfield  {journal} {\bibinfo  {journal} {Phys. Rev. X}\ }\textbf {\bibinfo {volume} {7}},\ \bibinfo {pages} {021026} (\bibinfo {year} {2017})}\BibitemShut {NoStop}%
\bibitem [{\citenamefont {Dimer}\ \emph {et~al.}(2007)\citenamefont {Dimer}, \citenamefont {Estienne}, \citenamefont {Parkins},\ and\ \citenamefont {Carmichael}}]{Dimer2007pro}%
  \BibitemOpen
  \bibfield  {author} {\bibinfo {author} {\bibfnamefont {F.}~\bibnamefont {Dimer}}, \bibinfo {author} {\bibfnamefont {B.}~\bibnamefont {Estienne}}, \bibinfo {author} {\bibfnamefont {A.~S.}\ \bibnamefont {Parkins}}, \ and\ \bibinfo {author} {\bibfnamefont {H.~J.}\ \bibnamefont {Carmichael}},\ }\bibfield  {title} {\enquote {\bibinfo {title} {{Proposed realization of the Dicke-model quantum phase transition in an optical cavity QED system}},}\ }\href {\doibase 10.1103/physreva.75.013804} {\bibfield  {journal} {\bibinfo  {journal} {Phys. Rev. A}\ }\textbf {\bibinfo {volume} {75}},\ \bibinfo {pages} {013804} (\bibinfo {year} {2007})}\BibitemShut {NoStop}%
\bibitem [{\citenamefont {Baumann}\ \emph {et~al.}(2010)\citenamefont {Baumann}, \citenamefont {Guerlin}, \citenamefont {Brennecke},\ and\ \citenamefont {Esslinger}}]{Baumann2010dqp}%
  \BibitemOpen
  \bibfield  {author} {\bibinfo {author} {\bibfnamefont {K.}~\bibnamefont {Baumann}}, \bibinfo {author} {\bibfnamefont {C.}~\bibnamefont {Guerlin}}, \bibinfo {author} {\bibfnamefont {F.}~\bibnamefont {Brennecke}}, \ and\ \bibinfo {author} {\bibfnamefont {T.}~\bibnamefont {Esslinger}},\ }\bibfield  {title} {\enquote {\bibinfo {title} {{Dicke quantum phase transition with a superfluid gas in an optical cavity}},}\ }\href {\doibase 10.1038/nature09009} {\bibfield  {journal} {\bibinfo  {journal} {Nature}\ }\textbf {\bibinfo {volume} {464}},\ \bibinfo {pages} {1301--1306} (\bibinfo {year} {2010})}\BibitemShut {NoStop}%
\bibitem [{\citenamefont {Kroeze}\ \emph {et~al.}(2018)\citenamefont {Kroeze}, \citenamefont {Guo}, \citenamefont {Vaidya}, \citenamefont {Keeling},\ and\ \citenamefont {Lev}}]{Kroeze2018sso}%
  \BibitemOpen
  \bibfield  {author} {\bibinfo {author} {\bibfnamefont {R.~M.}\ \bibnamefont {Kroeze}}, \bibinfo {author} {\bibfnamefont {Y.}~\bibnamefont {Guo}}, \bibinfo {author} {\bibfnamefont {V.~D.}\ \bibnamefont {Vaidya}}, \bibinfo {author} {\bibfnamefont {J.}~\bibnamefont {Keeling}}, \ and\ \bibinfo {author} {\bibfnamefont {B.~L.}\ \bibnamefont {Lev}},\ }\bibfield  {title} {\enquote {\bibinfo {title} {{Spinor Self-Ordering of a Quantum Gas in a Cavity}},}\ }\href {\doibase 10.1103/physrevlett.121.163601} {\bibfield  {journal} {\bibinfo  {journal} {Phys. Rev. Lett.}\ }\textbf {\bibinfo {volume} {121}},\ \bibinfo {pages} {163601} (\bibinfo {year} {2018})}\BibitemShut {NoStop}%
\bibitem [{\citenamefont {Carusotto}\ and\ \citenamefont {Ciuti}(2013)}]{Carusotto2013qfo}%
  \BibitemOpen
  \bibfield  {author} {\bibinfo {author} {\bibfnamefont {I.}~\bibnamefont {Carusotto}}\ and\ \bibinfo {author} {\bibfnamefont {C.}~\bibnamefont {Ciuti}},\ }\bibfield  {title} {\enquote {\bibinfo {title} {{Quantum fluids of light}},}\ }\href {\doibase 10.1103/revmodphys.85.299} {\bibfield  {journal} {\bibinfo  {journal} {Rev. Mod. Phys.}\ }\textbf {\bibinfo {volume} {85}},\ \bibinfo {pages} {299--366} (\bibinfo {year} {2013})}\BibitemShut {NoStop}%
\bibitem [{\citenamefont {Sanvitto}\ and\ \citenamefont {K\'ena-Cohen}(2016)}]{Sanvitto2016trt}%
  \BibitemOpen
  \bibfield  {author} {\bibinfo {author} {\bibfnamefont {D.}~\bibnamefont {Sanvitto}}\ and\ \bibinfo {author} {\bibfnamefont {S.}~\bibnamefont {K\'ena-Cohen}},\ }\bibfield  {title} {\enquote {\bibinfo {title} {{The road towards polaritonic devices}},}\ }\href {\doibase 10.1038/nmat4668} {\bibfield  {journal} {\bibinfo  {journal} {Nature Mater}\ }\textbf {\bibinfo {volume} {15}},\ \bibinfo {pages} {1061--1073} (\bibinfo {year} {2016})}\BibitemShut {NoStop}%
\bibitem [{\citenamefont {Schlawin}, \citenamefont {Cavalleri},\ and\ \citenamefont {Jaksch}(2019)}]{Schlawin2019ces}%
  \BibitemOpen
  \bibfield  {author} {\bibinfo {author} {\bibfnamefont {F.}~\bibnamefont {Schlawin}}, \bibinfo {author} {\bibfnamefont {A.}~\bibnamefont {Cavalleri}}, \ and\ \bibinfo {author} {\bibfnamefont {D.}~\bibnamefont {Jaksch}},\ }\bibfield  {title} {\enquote {\bibinfo {title} {{Cavity-Mediated Electron-Photon Superconductivity}},}\ }\href {\doibase 10.1103/physrevlett.122.133602} {\bibfield  {journal} {\bibinfo  {journal} {Phys. Rev. Lett.}\ }\textbf {\bibinfo {volume} {122}},\ \bibinfo {pages} {133602} (\bibinfo {year} {2019})}\BibitemShut {NoStop}%
\bibitem [{\citenamefont {Curtis}\ \emph {et~al.}(2019)\citenamefont {Curtis}, \citenamefont {Raines}, \citenamefont {Allocca}, \citenamefont {Hafezi},\ and\ \citenamefont {Galitski}}]{Curtis2019cqe}%
  \BibitemOpen
  \bibfield  {author} {\bibinfo {author} {\bibfnamefont {J.~B.}\ \bibnamefont {Curtis}}, \bibinfo {author} {\bibfnamefont {Z.~M.}\ \bibnamefont {Raines}}, \bibinfo {author} {\bibfnamefont {A.~A.}\ \bibnamefont {Allocca}}, \bibinfo {author} {\bibfnamefont {M.}~\bibnamefont {Hafezi}}, \ and\ \bibinfo {author} {\bibfnamefont {V.~M.}\ \bibnamefont {Galitski}},\ }\bibfield  {title} {\enquote {\bibinfo {title} {{Cavity Quantum Eliashberg Enhancement of Superconductivity}},}\ }\href {\doibase 10.1103/physrevlett.122.167002} {\bibfield  {journal} {\bibinfo  {journal} {Phys. Rev. Lett.}\ }\textbf {\bibinfo {volume} {122}},\ \bibinfo {pages} {167002} (\bibinfo {year} {2019})}\BibitemShut {NoStop}%
\bibitem [{\citenamefont {Gao}\ \emph {et~al.}(2020)\citenamefont {Gao}, \citenamefont {Schlawin}, \citenamefont {Buzzi}, \citenamefont {Cavalleri},\ and\ \citenamefont {Jaksch}}]{Gao2020pep}%
  \BibitemOpen
  \bibfield  {author} {\bibinfo {author} {\bibfnamefont {H.}~\bibnamefont {Gao}}, \bibinfo {author} {\bibfnamefont {F.}~\bibnamefont {Schlawin}}, \bibinfo {author} {\bibfnamefont {M.}~\bibnamefont {Buzzi}}, \bibinfo {author} {\bibfnamefont {A.}~\bibnamefont {Cavalleri}}, \ and\ \bibinfo {author} {\bibfnamefont {D.}~\bibnamefont {Jaksch}},\ }\bibfield  {title} {\enquote {\bibinfo {title} {{Photoinduced Electron Pairing in a Driven Cavity}},}\ }\href {\doibase 10.1103/physrevlett.125.053602} {\bibfield  {journal} {\bibinfo  {journal} {Phys. Rev. Lett.}\ }\textbf {\bibinfo {volume} {125}},\ \bibinfo {pages} {053602} (\bibinfo {year} {2020})}\BibitemShut {NoStop}%
\bibitem [{\citenamefont {Curtis}\ \emph {et~al.}(2022)\citenamefont {Curtis}, \citenamefont {Grankin}, \citenamefont {Poniatowski}, \citenamefont {Galitski}, \citenamefont {Narang},\ and\ \citenamefont {Demler}}]{Curtis2022cmi}%
  \BibitemOpen
  \bibfield  {author} {\bibinfo {author} {\bibfnamefont {J.~B.}\ \bibnamefont {Curtis}}, \bibinfo {author} {\bibfnamefont {A.}~\bibnamefont {Grankin}}, \bibinfo {author} {\bibfnamefont {N.~R.}\ \bibnamefont {Poniatowski}}, \bibinfo {author} {\bibfnamefont {V.~M.}\ \bibnamefont {Galitski}}, \bibinfo {author} {\bibfnamefont {P.}~\bibnamefont {Narang}}, \ and\ \bibinfo {author} {\bibfnamefont {E.}~\bibnamefont {Demler}},\ }\bibfield  {title} {\enquote {\bibinfo {title} {{Cavity magnon-polaritons in cuprate parent compounds}},}\ }\href {\doibase 10.1103/physrevresearch.4.013101} {\bibfield  {journal} {\bibinfo  {journal} {Phys. Rev. Research}\ }\textbf {\bibinfo {volume} {4}},\ \bibinfo {pages} {013101} (\bibinfo {year} {2022})}\BibitemShut {NoStop}%
\bibitem [{\citenamefont {Schlawin}, \citenamefont {Kennes},\ and\ \citenamefont {Sentef}(2022)}]{Schlawin2022cqm}%
  \BibitemOpen
  \bibfield  {author} {\bibinfo {author} {\bibfnamefont {F.}~\bibnamefont {Schlawin}}, \bibinfo {author} {\bibfnamefont {D.~M.}\ \bibnamefont {Kennes}}, \ and\ \bibinfo {author} {\bibfnamefont {M.~A.}\ \bibnamefont {Sentef}},\ }\bibfield  {title} {\enquote {\bibinfo {title} {{Cavity quantum materials}},}\ }\href {\doibase 10.1063/5.0083825} {\bibfield  {journal} {\bibinfo  {journal} {Applied Physics Reviews}\ }\textbf {\bibinfo {volume} {9}},\ \bibinfo {pages} {011312} (\bibinfo {year} {2022})}\BibitemShut {NoStop}%
\bibitem [{\citenamefont {Bretscher}\ \emph {et~al.}(2026)\citenamefont {Bretscher}, \citenamefont {Graziotto}, \citenamefont {Michael}, \citenamefont {Montanaro}, \citenamefont {Lu}, \citenamefont {Grankin}, \citenamefont {McIver}, \citenamefont {Faist}, \citenamefont {Fausti}, \citenamefont {Eckstein}, \citenamefont {Ruggenthaler}, \citenamefont {Rubio}, \citenamefont {Basov}, \citenamefont {Hafezi}, \citenamefont {Claassen}, \citenamefont {Kennes},\ and\ \citenamefont {Sentef}}]{Bretscher2026fei}%
  \BibitemOpen
  \bibfield  {author} {\bibinfo {author} {\bibfnamefont {H.~M.}\ \bibnamefont {Bretscher}}, \bibinfo {author} {\bibfnamefont {L.}~\bibnamefont {Graziotto}}, \bibinfo {author} {\bibfnamefont {M.~H.}\ \bibnamefont {Michael}}, \bibinfo {author} {\bibfnamefont {A.}~\bibnamefont {Montanaro}}, \bibinfo {author} {\bibfnamefont {I.-T.}\ \bibnamefont {Lu}}, \bibinfo {author} {\bibfnamefont {A.}~\bibnamefont {Grankin}}, \bibinfo {author} {\bibfnamefont {J.~W.}\ \bibnamefont {McIver}}, \bibinfo {author} {\bibfnamefont {J.}~\bibnamefont {Faist}}, \bibinfo {author} {\bibfnamefont {D.}~\bibnamefont {Fausti}}, \bibinfo {author} {\bibfnamefont {M.}~\bibnamefont {Eckstein}}, \bibinfo {author} {\bibfnamefont {M.}~\bibnamefont {Ruggenthaler}}, \bibinfo {author} {\bibfnamefont {A.}~\bibnamefont {Rubio}}, \bibinfo {author} {\bibfnamefont {D.}~\bibnamefont {Basov}}, \bibinfo {author} {\bibfnamefont {M.}~\bibnamefont {Hafezi}}, \bibinfo {author} {\bibfnamefont {M.}~\bibnamefont {Claassen}}, \bibinfo {author} {\bibfnamefont {D.~M.}\
  \bibnamefont {Kennes}}, \ and\ \bibinfo {author} {\bibfnamefont {M.~A.}\ \bibnamefont {Sentef}},\ }\bibfield  {title} {\enquote {\bibinfo {title} {{Fluctuation engineering in cavity quantum materials}},}\ }\href {https://arxiv.org/abs/2604.08666} {\  (\bibinfo {year} {2026})},\ \Eprint {http://arxiv.org/abs/arXiv:2604.08666} {arXiv:2604.08666} \BibitemShut {NoStop}%
\bibitem [{\citenamefont {Kirton}\ and\ \citenamefont {Keeling}(2018)}]{Kirton2018sal}%
  \BibitemOpen
  \bibfield  {author} {\bibinfo {author} {\bibfnamefont {P.}~\bibnamefont {Kirton}}\ and\ \bibinfo {author} {\bibfnamefont {J.}~\bibnamefont {Keeling}},\ }\bibfield  {title} {\enquote {\bibinfo {title} {{Superradiant and lasing states in driven-dissipative Dicke models}},}\ }\href {\doibase 10.1088/1367-2630/aaa11d} {\bibfield  {journal} {\bibinfo  {journal} {New J. Phys.}\ }\textbf {\bibinfo {volume} {20}},\ \bibinfo {pages} {015009} (\bibinfo {year} {2018})}\BibitemShut {NoStop}%
\bibitem [{\citenamefont {Kimble}(1998)}]{Kimble1998sio}%
  \BibitemOpen
  \bibfield  {author} {\bibinfo {author} {\bibfnamefont {H.~J.}\ \bibnamefont {Kimble}},\ }\bibfield  {title} {\enquote {\bibinfo {title} {{Strong Interactions of Single Atoms and Photons in Cavity QED}},}\ }\href {\doibase 10.1238/physica.topical.076a00127} {\bibfield  {journal} {\bibinfo  {journal} {Phys. Scr.}\ }\textbf {\bibinfo {volume} {T76}},\ \bibinfo {pages} {127} (\bibinfo {year} {1998})}\BibitemShut {NoStop}%
\bibitem [{\citenamefont {Liu}\ \emph {et~al.}(2015)\citenamefont {Liu}, \citenamefont {Galfsky}, \citenamefont {Sun}, \citenamefont {Xia}, \citenamefont {Lin}, \citenamefont {Lee}, \citenamefont {K\'{e}na-Cohen},\ and\ \citenamefont {Menon}}]{Liu2015}%
  \BibitemOpen
  \bibfield  {author} {\bibinfo {author} {\bibfnamefont {X.}~\bibnamefont {Liu}}, \bibinfo {author} {\bibfnamefont {T.}~\bibnamefont {Galfsky}}, \bibinfo {author} {\bibfnamefont {Z.}~\bibnamefont {Sun}}, \bibinfo {author} {\bibfnamefont {F.}~\bibnamefont {Xia}}, \bibinfo {author} {\bibfnamefont {E.-c.}\ \bibnamefont {Lin}}, \bibinfo {author} {\bibfnamefont {Y.-H.}\ \bibnamefont {Lee}}, \bibinfo {author} {\bibfnamefont {S.}~\bibnamefont {K\'{e}na-Cohen}}, \ and\ \bibinfo {author} {\bibfnamefont {V.~M.}\ \bibnamefont {Menon}},\ }\bibfield  {title} {\enquote {\bibinfo {title} {{Strong light-matter coupling in two-dimensional atomic crystals}},}\ }\href {\doibase 10.1038/nphoton.2014.304} {\bibfield  {journal} {\bibinfo  {journal} {Nat. Photonics}\ }\textbf {\bibinfo {volume} {9}},\ \bibinfo {pages} {30--34} (\bibinfo {year} {2015})}\BibitemShut {NoStop}%
\bibitem [{\citenamefont {Dufferwiel}\ \emph {et~al.}(2015)\citenamefont {Dufferwiel}, \citenamefont {Schwarz}, \citenamefont {Withers}, \citenamefont {Trichet}, \citenamefont {Li}, \citenamefont {Sich}, \citenamefont {Del Pozo-Zamudio}, \citenamefont {Clark}, \citenamefont {Nalitov}, \citenamefont {Solnyshkov}, \citenamefont {Malpuech}, \citenamefont {Novoselov}, \citenamefont {Smith}, \citenamefont {Skolnick}, \citenamefont {Krizhanovskii},\ and\ \citenamefont {Tartakovskii}}]{Dufferwiel2015}%
  \BibitemOpen
  \bibfield  {author} {\bibinfo {author} {\bibfnamefont {S.}~\bibnamefont {Dufferwiel}}, \bibinfo {author} {\bibfnamefont {S.}~\bibnamefont {Schwarz}}, \bibinfo {author} {\bibfnamefont {F.}~\bibnamefont {Withers}}, \bibinfo {author} {\bibfnamefont {A.~A.~P.}\ \bibnamefont {Trichet}}, \bibinfo {author} {\bibfnamefont {F.}~\bibnamefont {Li}}, \bibinfo {author} {\bibfnamefont {M.}~\bibnamefont {Sich}}, \bibinfo {author} {\bibfnamefont {O.}~\bibnamefont {Del Pozo-Zamudio}}, \bibinfo {author} {\bibfnamefont {C.}~\bibnamefont {Clark}}, \bibinfo {author} {\bibfnamefont {A.}~\bibnamefont {Nalitov}}, \bibinfo {author} {\bibfnamefont {D.~D.}\ \bibnamefont {Solnyshkov}}, \bibinfo {author} {\bibfnamefont {G.}~\bibnamefont {Malpuech}}, \bibinfo {author} {\bibfnamefont {K.~S.}\ \bibnamefont {Novoselov}}, \bibinfo {author} {\bibfnamefont {J.~M.}\ \bibnamefont {Smith}}, \bibinfo {author} {\bibfnamefont {M.~S.}\ \bibnamefont {Skolnick}}, \bibinfo {author} {\bibfnamefont {D.~N.}\ \bibnamefont {Krizhanovskii}}, \ and\ \bibinfo
  {author} {\bibfnamefont {A.~I.}\ \bibnamefont {Tartakovskii}},\ }\bibfield  {title} {\enquote {\bibinfo {title} {{Exciton-polaritons in van der Waals heterostructures embedded in tunable microcavities}},}\ }\href {\doibase 10.1038/ncomms9579} {\bibfield  {journal} {\bibinfo  {journal} {Nat. Commun.}\ }\textbf {\bibinfo {volume} {6}},\ \bibinfo {pages} {8579} (\bibinfo {year} {2015})}\BibitemShut {NoStop}%
\bibitem [{\citenamefont {Zhang}\ \emph {et~al.}(2021)\citenamefont {Zhang}, \citenamefont {Wu}, \citenamefont {Hou}, \citenamefont {Zhang}, \citenamefont {Chou}, \citenamefont {Watanabe}, \citenamefont {Taniguchi}, \citenamefont {Forrest},\ and\ \citenamefont {Deng}}]{Zhang2021vdw}%
  \BibitemOpen
  \bibfield  {author} {\bibinfo {author} {\bibfnamefont {L.}~\bibnamefont {Zhang}}, \bibinfo {author} {\bibfnamefont {F.}~\bibnamefont {Wu}}, \bibinfo {author} {\bibfnamefont {S.}~\bibnamefont {Hou}}, \bibinfo {author} {\bibfnamefont {Z.}~\bibnamefont {Zhang}}, \bibinfo {author} {\bibfnamefont {Y.-H.}\ \bibnamefont {Chou}}, \bibinfo {author} {\bibfnamefont {K.}~\bibnamefont {Watanabe}}, \bibinfo {author} {\bibfnamefont {T.}~\bibnamefont {Taniguchi}}, \bibinfo {author} {\bibfnamefont {S.~R.}\ \bibnamefont {Forrest}}, \ and\ \bibinfo {author} {\bibfnamefont {H.}~\bibnamefont {Deng}},\ }\bibfield  {title} {\enquote {\bibinfo {title} {{Van der Waals heterostructure polaritons with moiré-induced nonlinearity}},}\ }\href {\doibase 10.1038/s41586-021-03228-5} {\bibfield  {journal} {\bibinfo  {journal} {Nature}\ }\textbf {\bibinfo {volume} {591}},\ \bibinfo {pages} {61--65} (\bibinfo {year} {2021})}\BibitemShut {NoStop}%
\bibitem [{\citenamefont {Vadia}\ \emph {et~al.}(2021)\citenamefont {Vadia}, \citenamefont {Scherzer}, \citenamefont {Thierschmann}, \citenamefont {Sch\"afermeier}, \citenamefont {Dal~Savio}, \citenamefont {Taniguchi}, \citenamefont {Watanabe}, \citenamefont {Hunger}, \citenamefont {Karra\"\i},\ and\ \citenamefont {H\"ogele}}]{Vadia2021oic}%
  \BibitemOpen
  \bibfield  {author} {\bibinfo {author} {\bibfnamefont {S.}~\bibnamefont {Vadia}}, \bibinfo {author} {\bibfnamefont {J.}~\bibnamefont {Scherzer}}, \bibinfo {author} {\bibfnamefont {H.}~\bibnamefont {Thierschmann}}, \bibinfo {author} {\bibfnamefont {C.}~\bibnamefont {Sch\"afermeier}}, \bibinfo {author} {\bibfnamefont {C.}~\bibnamefont {Dal~Savio}}, \bibinfo {author} {\bibfnamefont {T.}~\bibnamefont {Taniguchi}}, \bibinfo {author} {\bibfnamefont {K.}~\bibnamefont {Watanabe}}, \bibinfo {author} {\bibfnamefont {D.}~\bibnamefont {Hunger}}, \bibinfo {author} {\bibfnamefont {K.}~\bibnamefont {Karra\"\i}}, \ and\ \bibinfo {author} {\bibfnamefont {A.}~\bibnamefont {H\"ogele}},\ }\bibfield  {title} {\enquote {\bibinfo {title} {{Open-Cavity in Closed-Cycle Cryostat as a Quantum Optics Platform}},}\ }\href {\doibase 10.1103/prxquantum.2.040318} {\bibfield  {journal} {\bibinfo  {journal} {PRX Quantum}\ }\textbf {\bibinfo {volume} {2}},\ \bibinfo {pages} {040318} (\bibinfo {year} {2021})}\BibitemShut {NoStop}%
\bibitem [{\citenamefont {Drawer}\ \emph {et~al.}(2025)\citenamefont {Drawer}, \citenamefont {Cianci}, \citenamefont {Solovyeva}, \citenamefont {Steinhoff}, \citenamefont {Gies}, \citenamefont {Eilenberger}, \citenamefont {Watanabe}, \citenamefont {Taniguchi}, \citenamefont {Solovev}, \citenamefont {Pettinari}, \citenamefont {Tuzi}, \citenamefont {Blundo}, \citenamefont {Felici}, \citenamefont {Polimeni}, \citenamefont {Esmann},\ and\ \citenamefont {Schneider}}]{Drawer2025twm}%
  \BibitemOpen
  \bibfield  {author} {\bibinfo {author} {\bibfnamefont {J.-C.}\ \bibnamefont {Drawer}}, \bibinfo {author} {\bibfnamefont {S.}~\bibnamefont {Cianci}}, \bibinfo {author} {\bibfnamefont {V.}~\bibnamefont {Solovyeva}}, \bibinfo {author} {\bibfnamefont {A.}~\bibnamefont {Steinhoff}}, \bibinfo {author} {\bibfnamefont {C.}~\bibnamefont {Gies}}, \bibinfo {author} {\bibfnamefont {F.}~\bibnamefont {Eilenberger}}, \bibinfo {author} {\bibfnamefont {K.}~\bibnamefont {Watanabe}}, \bibinfo {author} {\bibfnamefont {T.}~\bibnamefont {Taniguchi}}, \bibinfo {author} {\bibfnamefont {I.}~\bibnamefont {Solovev}}, \bibinfo {author} {\bibfnamefont {G.}~\bibnamefont {Pettinari}}, \bibinfo {author} {\bibfnamefont {F.}~\bibnamefont {Tuzi}}, \bibinfo {author} {\bibfnamefont {E.}~\bibnamefont {Blundo}}, \bibinfo {author} {\bibfnamefont {M.}~\bibnamefont {Felici}}, \bibinfo {author} {\bibfnamefont {A.}~\bibnamefont {Polimeni}}, \bibinfo {author} {\bibfnamefont {M.}~\bibnamefont {Esmann}}, \ and\ \bibinfo {author} {\bibfnamefont
  {C.}~\bibnamefont {Schneider}},\ }\bibfield  {title} {\enquote {\bibinfo {title} {{Tunable WS$_2$ Micro-Dome Open Cavity Single Photon Source}},}\ }\href {https://arxiv.org/abs/2511.21630} {\  (\bibinfo {year} {2025})},\ \Eprint {http://arxiv.org/abs/arXiv:2511.21630} {arXiv:2511.21630} \BibitemShut {NoStop}%
\bibitem [{\citenamefont {Hoang}\ \emph {et~al.}(2026)\citenamefont {Hoang}, \citenamefont {Mahdikhany}, \citenamefont {Wang}, \citenamefont {Mirin}, \citenamefont {Silverman}, \citenamefont {Imany},\ and\ \citenamefont {Sun}}]{Hoang2026acr}%
  \BibitemOpen
  \bibfield  {author} {\bibinfo {author} {\bibfnamefont {T.~D.}\ \bibnamefont {Hoang}}, \bibinfo {author} {\bibfnamefont {F.}~\bibnamefont {Mahdikhany}}, \bibinfo {author} {\bibfnamefont {Z.}~\bibnamefont {Wang}}, \bibinfo {author} {\bibfnamefont {R.}~\bibnamefont {Mirin}}, \bibinfo {author} {\bibfnamefont {K.}~\bibnamefont {Silverman}}, \bibinfo {author} {\bibfnamefont {P.}~\bibnamefont {Imany}}, \ and\ \bibinfo {author} {\bibfnamefont {S.}~\bibnamefont {Sun}},\ }\bibfield  {title} {\enquote {\bibinfo {title} {{A Compact, Robust, and Tunable Open Microcavity Platform for Solid-State Quantum Electrodynamics}},}\ }\href {https://arxiv.org/abs/2605.16694} {\  (\bibinfo {year} {2026})},\ \Eprint {http://arxiv.org/abs/arXiv:2605.16694} {arXiv:2605.16694} \BibitemShut {NoStop}%
\bibitem [{\citenamefont {Vaidya}\ \emph {et~al.}(2018)\citenamefont {Vaidya}, \citenamefont {Guo}, \citenamefont {Kroeze}, \citenamefont {Ballantine}, \citenamefont {Koll\'{a}r}, \citenamefont {Keeling},\ and\ \citenamefont {Lev}}]{Vaidya2018tpa}%
  \BibitemOpen
  \bibfield  {author} {\bibinfo {author} {\bibfnamefont {V.~D.}\ \bibnamefont {Vaidya}}, \bibinfo {author} {\bibfnamefont {Y.}~\bibnamefont {Guo}}, \bibinfo {author} {\bibfnamefont {R.~M.}\ \bibnamefont {Kroeze}}, \bibinfo {author} {\bibfnamefont {K.~E.}\ \bibnamefont {Ballantine}}, \bibinfo {author} {\bibfnamefont {A.~J.}\ \bibnamefont {Koll\'{a}r}}, \bibinfo {author} {\bibfnamefont {J.}~\bibnamefont {Keeling}}, \ and\ \bibinfo {author} {\bibfnamefont {B.~L.}\ \bibnamefont {Lev}},\ }\bibfield  {title} {\enquote {\bibinfo {title} {{Tunable-Range, Photon-Mediated Atomic Interactions in Multimode Cavity QED}},}\ }\href {\doibase 10.1103/physrevx.8.011002} {\bibfield  {journal} {\bibinfo  {journal} {Phys. Rev. X}\ }\textbf {\bibinfo {volume} {8}},\ \bibinfo {pages} {011002} (\bibinfo {year} {2018})}\BibitemShut {NoStop}%
\bibitem [{\citenamefont {Guo}\ \emph {et~al.}(2019{\natexlab{a}})\citenamefont {Guo}, \citenamefont {Kroeze}, \citenamefont {Vaidya}, \citenamefont {Keeling},\ and\ \citenamefont {Lev}}]{Guo2019spa}%
  \BibitemOpen
  \bibfield  {author} {\bibinfo {author} {\bibfnamefont {Y.}~\bibnamefont {Guo}}, \bibinfo {author} {\bibfnamefont {R.~M.}\ \bibnamefont {Kroeze}}, \bibinfo {author} {\bibfnamefont {V.~D.}\ \bibnamefont {Vaidya}}, \bibinfo {author} {\bibfnamefont {J.}~\bibnamefont {Keeling}}, \ and\ \bibinfo {author} {\bibfnamefont {B.~L.}\ \bibnamefont {Lev}},\ }\bibfield  {title} {\enquote {\bibinfo {title} {{Sign-Changing Photon-Mediated Atom Interactions in Multimode Cavity Quantum Electrodynamics}},}\ }\href {\doibase 10.1103/physrevlett.122.193601} {\bibfield  {journal} {\bibinfo  {journal} {Phys. Rev. Lett.}\ }\textbf {\bibinfo {volume} {122}},\ \bibinfo {pages} {193601} (\bibinfo {year} {2019}{\natexlab{a}})}\BibitemShut {NoStop}%
\bibitem [{\citenamefont {Guo}\ \emph {et~al.}(2019{\natexlab{b}})\citenamefont {Guo}, \citenamefont {Vaidya}, \citenamefont {Kroeze}, \citenamefont {Lunney}, \citenamefont {Lev},\ and\ \citenamefont {Keeling}}]{Guo2019eab}%
  \BibitemOpen
  \bibfield  {author} {\bibinfo {author} {\bibfnamefont {Y.}~\bibnamefont {Guo}}, \bibinfo {author} {\bibfnamefont {V.~D.}\ \bibnamefont {Vaidya}}, \bibinfo {author} {\bibfnamefont {R.~M.}\ \bibnamefont {Kroeze}}, \bibinfo {author} {\bibfnamefont {R.~A.}\ \bibnamefont {Lunney}}, \bibinfo {author} {\bibfnamefont {B.~L.}\ \bibnamefont {Lev}}, \ and\ \bibinfo {author} {\bibfnamefont {J.}~\bibnamefont {Keeling}},\ }\bibfield  {title} {\enquote {\bibinfo {title} {{Emergent and broken symmetries of atomic self-organization arising from Gouy phase shifts in multimode cavity QED}},}\ }\href {\doibase 10.1103/physreva.99.053818} {\bibfield  {journal} {\bibinfo  {journal} {Phys. Rev. A}\ }\textbf {\bibinfo {volume} {99}},\ \bibinfo {pages} {053818} (\bibinfo {year} {2019}{\natexlab{b}})}\BibitemShut {NoStop}%
\bibitem [{\citenamefont {Kroeze}\ \emph {et~al.}(2023)\citenamefont {Kroeze}, \citenamefont {Marsh}, \citenamefont {Lin}, \citenamefont {Keeling},\ and\ \citenamefont {Lev}}]{Kroeze2023hcu}%
  \BibitemOpen
  \bibfield  {author} {\bibinfo {author} {\bibfnamefont {R.~M.}\ \bibnamefont {Kroeze}}, \bibinfo {author} {\bibfnamefont {B.~P.}\ \bibnamefont {Marsh}}, \bibinfo {author} {\bibfnamefont {K.-Y.}\ \bibnamefont {Lin}}, \bibinfo {author} {\bibfnamefont {J.}~\bibnamefont {Keeling}}, \ and\ \bibinfo {author} {\bibfnamefont {B.~L.}\ \bibnamefont {Lev}},\ }\bibfield  {title} {\enquote {\bibinfo {title} {{High Cooperativity Using a Confocal-Cavity–QED Microscope}},}\ }\href {\doibase 10.1103/prxquantum.4.020326} {\bibfield  {journal} {\bibinfo  {journal} {PRX Quantum}\ }\textbf {\bibinfo {volume} {4}},\ \bibinfo {pages} {020326} (\bibinfo {year} {2023})}\BibitemShut {NoStop}%
\bibitem [{\citenamefont {Skolc}\ \emph {et~al.}(2026)\citenamefont {Skolc}, \citenamefont {Chattopadhyay}, \citenamefont {Marijanović}, \citenamefont {Li}, \citenamefont {Keeling}, \citenamefont {Lev},\ and\ \citenamefont {Demler}}]{Skolc2026scd}%
  \BibitemOpen
  \bibfield  {author} {\bibinfo {author} {\bibfnamefont {L.}~\bibnamefont {Skolc}}, \bibinfo {author} {\bibfnamefont {S.}~\bibnamefont {Chattopadhyay}}, \bibinfo {author} {\bibfnamefont {F.}~\bibnamefont {Marijanović}}, \bibinfo {author} {\bibfnamefont {Q.}~\bibnamefont {Li}}, \bibinfo {author} {\bibfnamefont {J.}~\bibnamefont {Keeling}}, \bibinfo {author} {\bibfnamefont {B.~L.}\ \bibnamefont {Lev}}, \ and\ \bibinfo {author} {\bibfnamefont {E.}~\bibnamefont {Demler}},\ }\bibfield  {title} {\enquote {\bibinfo {title} {{Superradiant Charge Density Waves in a Driven Cavity-Matter Hybrid}},}\ }\href {\doibase 10.48550/arxiv.2603.28432} {\  (\bibinfo {year} {2026}),\ 10.48550/arxiv.2603.28432},\ \Eprint {http://arxiv.org/abs/arXiv:2603.28432} {arXiv:2603.28432} \BibitemShut {NoStop}%
\bibitem [{\citenamefont {Naides}\ \emph {et~al.}(2013)\citenamefont {Naides}, \citenamefont {Turner}, \citenamefont {Lai}, \citenamefont {DiSciacca},\ and\ \citenamefont {Lev}}]{Naides2013tug}%
  \BibitemOpen
  \bibfield  {author} {\bibinfo {author} {\bibfnamefont {M.~A.}\ \bibnamefont {Naides}}, \bibinfo {author} {\bibfnamefont {R.~W.}\ \bibnamefont {Turner}}, \bibinfo {author} {\bibfnamefont {R.~A.}\ \bibnamefont {Lai}}, \bibinfo {author} {\bibfnamefont {J.~M.}\ \bibnamefont {DiSciacca}}, \ and\ \bibinfo {author} {\bibfnamefont {B.~L.}\ \bibnamefont {Lev}},\ }\bibfield  {title} {\enquote {\bibinfo {title} {{Trapping ultracold gases near cryogenic materials with rapid reconfigurability}},}\ }\href {\doibase 10.1063/1.4852017} {\bibfield  {journal} {\bibinfo  {journal} {Appl. Phys. Lett.}\ }\textbf {\bibinfo {volume} {103}},\ \bibinfo {pages} {251112} (\bibinfo {year} {2013})}\BibitemShut {NoStop}%
\bibitem [{\citenamefont {Koll\'ar}\ \emph {et~al.}(2015)\citenamefont {Koll\'ar}, \citenamefont {Papageorge}, \citenamefont {Baumann}, \citenamefont {Armen},\ and\ \citenamefont {Lev}}]{Kollar2015aac}%
  \BibitemOpen
  \bibfield  {author} {\bibinfo {author} {\bibfnamefont {A.~J.}\ \bibnamefont {Koll\'ar}}, \bibinfo {author} {\bibfnamefont {A.~T.}\ \bibnamefont {Papageorge}}, \bibinfo {author} {\bibfnamefont {K.}~\bibnamefont {Baumann}}, \bibinfo {author} {\bibfnamefont {M.~A.}\ \bibnamefont {Armen}}, \ and\ \bibinfo {author} {\bibfnamefont {B.~L.}\ \bibnamefont {Lev}},\ }\bibfield  {title} {\enquote {\bibinfo {title} {{An adjustable-length cavity and Bose–Einstein condensate apparatus for multimode cavity QED}},}\ }\href {\doibase 10.1088/1367-2630/17/4/043012} {\bibfield  {journal} {\bibinfo  {journal} {New J. Phys.}\ }\textbf {\bibinfo {volume} {17}},\ \bibinfo {pages} {043012} (\bibinfo {year} {2015})}\BibitemShut {NoStop}%
\bibitem [{\citenamefont {Taylor}\ \emph {et~al.}(2021)\citenamefont {Taylor}, \citenamefont {Yang}, \citenamefont {Freudenstein},\ and\ \citenamefont {Lev}}]{Taylor2021asq}%
  \BibitemOpen
  \bibfield  {author} {\bibinfo {author} {\bibfnamefont {S.}~\bibnamefont {Taylor}}, \bibinfo {author} {\bibfnamefont {F.}~\bibnamefont {Yang}}, \bibinfo {author} {\bibfnamefont {B.~A.}\ \bibnamefont {Freudenstein}}, \ and\ \bibinfo {author} {\bibfnamefont {B.}~\bibnamefont {Lev}},\ }\bibfield  {title} {\enquote {\bibinfo {title} {{A scanning quantum cryogenic atom microscope at 6 K}},}\ }\href {\doibase 10.21468/scipostphys.10.3.060} {\bibfield  {journal} {\bibinfo  {journal} {SciPost Phys.}\ }\textbf {\bibinfo {volume} {10}},\ \bibinfo {pages} {60} (\bibinfo {year} {2021})}\BibitemShut {NoStop}%
\bibitem [{\citenamefont {Shirley}(1982)}]{Shirley1982}%
  \BibitemOpen
  \bibfield  {author} {\bibinfo {author} {\bibfnamefont {J.~H.}\ \bibnamefont {Shirley}},\ }\bibfield  {title} {\enquote {\bibinfo {title} {Modulation transfer processes in optical heterodyne saturation spectroscopy},}\ }\href {\doibase 10.1364/OL.7.000537} {\bibfield  {journal} {\bibinfo  {journal} {Opt. Lett.}\ }\textbf {\bibinfo {volume} {7}},\ \bibinfo {pages} {537--539} (\bibinfo {year} {1982})}\BibitemShut {NoStop}%
\bibitem [{\citenamefont {Black}(2001)}]{Black2001ait}%
  \BibitemOpen
  \bibfield  {author} {\bibinfo {author} {\bibfnamefont {E.~D.}\ \bibnamefont {Black}},\ }\bibfield  {title} {\enquote {\bibinfo {title} {{An introduction to Pound–Drever–Hall laser frequency stabilization}},}\ }\href {\doibase 10.1119/1.1286663} {\bibfield  {journal} {\bibinfo  {journal} {Am. J. Phys}\ }\textbf {\bibinfo {volume} {69}},\ \bibinfo {pages} {79--87} (\bibinfo {year} {2001})}\BibitemShut {NoStop}%
\end{thebibliography}
\end{document}